\documentclass[a4paper,fleqn,usenatbib,useAMS]{mnras}
\usepackage{graphics}
\usepackage{graphicx}
\usepackage{amsmath}    % Advanced maths commands
\usepackage{amssymb}    % Extra maths symbols
\usepackage{multicol}        % Multi-column entries in tables
\usepackage{bm}     % Bold maths symbols, including upright Greek
\usepackage{pdflscape}  % Landscape pages
\usepackage{xcolor}
\usepackage{ae,aecompl}
\usepackage{subfigure}
\usepackage{scalerel}
\usepackage[english]{babel}
\usepackage{times}
\usepackage{physics}
\usepackage{etoolbox}
\usepackage{multirow, booktabs}
\usepackage{siunitx}
\usepackage{paralist}
\usepackage{makecell}
\usepackage{tabularx}
\usepackage[inline]{enumitem}
\newlist{mycompactenum}{enumerate}{1}
\setlist[mycompactenum,1]{nosep,label=\arabic*.}
\usepackage{paralist}

\newcommand{\wfirst}{\textit{Roman}}

\title[Microlensing due to FFPs towards the MCs]{On the detection of free-floating planets through microlensing towards the Magellanic Clouds}
\author[Sajadian]{Sedighe Sajadian$~^{1,2}$\thanks{E-mail: s.sajadian@iut.ac.ir}\\
	$^{1}$Department~of~Physics,~Isfahan~University~of~Technology,~Isfahan~84156-83111,~Iran\\
	$^{2}$Department~of~Physics,~Chungbuk~National~University,~Cheongju~28644,~Republic~of~Korea}

\begin{document}
\label{firstpage}
\pagerange{\pageref{firstpage}--\pageref{lastpage}}
\maketitle
\begin{abstract}
In this work, we study detecting free-floating planets (FFPs) by microlensing observations towards the Magellanic Clouds (MCs). In comparison to similar events toward the Galactic bulge, an FFP in the Galactic halo produces on average longer microlensing events with smaller projected source radii toward these clouds. For these microlensing events, the relative lens-source velocities are on average smaller. The MC self-lensing events due to FFPs suffer from severe finite-source effects. We first simulate microlensing events due to FFPs towards MCs and assume a log-uniform step function for their mass. The efficiencies for capturing their lensing signatures (with signal-to-noise greater than $50$) are found to be $0.1$-$0.6\%$ and $3$-$6\%$ through ground-based optical surveys and space-based near-infrared surveys, respectively. We then promote these simulations and assume the \wfirst~telescope continuously observes each MC during one $72$-day season with the $15$min observing cadence. From simulated microlensing events with the resolvable source stars at the baseline due to FFPs with the masses $\sim 0.01$-$10^{4}M_{\oplus}$, \wfirst~discovers their lensing effects with the efficiencies $\sim 10$-$80\%$, respectively. By adopting $5\%$ as halos fraction from FFPs we estimate the expected number of events. The highest number of detectable FFPs which is $\sim1700$-$2200$ per season per square degree happens for ones with masses $\sim 0.01 M_{\oplus}$. Our simulations show that \wfirst~potentially extends the mass range of detectable FFPs in halos down to $5.9\times 10^{-7} M_{\oplus}$ with continuous observations during one observing season from the Large Magellanic Cloud.
\end{abstract}


%%%%%%%%%%%%%%%%%%%%%%%%%%%%%%%%%%%%%%%%%%
\section{Introduction}
%%%%%%%%% Gravitational microlensing and free floatign exoplanets 
Free-floating or rogue planets (FFPs) are isolated objects that do not orbit a parent star. They have masses similar to those of planets. Two main mechanisms have been introduced to generate low-mass FFPs. (i) Dynamical instability which causes planet-planet scatterings and, as a result, the ejection of one or more planet(s) from planetary systems \citep{2006Boss, Veras2012}. This method is more efficient for ejecting Earth- and super Earth-mass planets from planetary systems rather than giant and Jupiter-like planets \citep{2015Pfyffer, Coreaccretion, 2017Barclay}. (ii) Core accretion theory of planet formation in protoplanetary discs which can generate both bound planets and FFPs \citep{IdaLin2004, 2013IdaLin,2009Mordasini}.

Up to now, several massive and young FFPs or brown dwarfs in young star-forming regions have been detected through deep direct imaging in infrared passband or wide-field surveys \citep[see, e.g.,][]{Bihain2009, Delorme2012, Liu2013, Dupuy2013, 2017Gagne, Rodriguez2017, Bejar2018, 2019Bardalez}. Low-mass planets ($\lesssim 5M_{\rm J}$, where $M_{\rm J}$ is the mass of Jupiter) which are dark have no chance to be detected by these methods. These objects can be discerned through their gravitational effects on the light path of a co-linear and background source star \citep{Hanetal2004,DiStefano2012,Bennett2012}.

The gravitational effect due to a massive object on the light path of a background source star which is a result of Einstein's theory of general relativity  makes two or more distorted images from that star \citep{Einstein1936, Liebes1964,ChangNatur}. In the Galactic scale, the angular distance of these images is too small to be resolved even by space-based telescopes and an observer receives the total light of both images which is magnified. This effect is the so-called gravitational microlensing \citep[see, e.g., ][]{schneider1992, Gaudi2012, Mao2012}. The characteristic timescale of a microlensing event is proportional to the square root of the lens mass, i.e., $t_{\rm E}\propto \sqrt{M_{\rm l}}$, where $M_{\rm l}$ is the lens mass. If an FFP acts as the lens object, the resulting timescale is short and mostly less than $2~$days for microlensing events towards the Galactic bulge. These events are the so-called short-duration microlensing events \citep[see, e.g.,][]{Hanetal.2005}. However, short-duration microlensing events, in a degenerate way, are generated when relative lens-source velocities are high or lens objects are very close to the observer \citep[see, e.g.,][]{DiStefano2012,2011NaturWambsganss}.

The Microlensing Observations in Astrophysics (MOA, \citet{MOA_gourp}) microlensing group in 2011 has detected an excess of short-duration microlensing events due to FFPs towards the Galactic bulge \citep{Sumi2011Natur}. The next observation was done by the Optical Gravitational Lensing Experiment (OGLE, \citet{OGLE_IV}) group in 2017 and they found a lower number of short-duration microlensing events \citep{Mroz2017Natur}. Furthermore, several short-duration microlensing events due to FFPs have been discovered by the OGLE and Korea Microlensing Telescope Network (KMTNet, \citet{KMTNet2016}) surveys \citep{2018Mroz_1, 2019Mroz_1, 2020Mroz_1, 2020Kim_1,2020Han_1, 2021Ryu_1}. We note that two previously mentioned mechanisms for generating FFPs were unable to generate the number of FFPs predicted by the MOA-II observation \citep{Coreaccretion, 2011NaturWambsganss, 2012MNRASVeras}.

In order to find a reasonable statistic of FFPs and examine previous observational results, the upcoming survey observation, \textit{The Nancy Grace Roman Space Telescope} (\wfirst) survey, will search for FFPs and short-duration microlensing events towards the Galactic bulge during 6 $72$-day observing seasons with improved cadence, i.e., $\sim15$ min \citep{Penny2019,Penney2020}. 

\noindent Additionally, FFPs in the Galactic halo can be probed through survey observations of microlensing events towards nearby galaxies, e.g., the Magellanic Clouds, M31, etc. \citep{Paczynski1986}. Searching for these microlensing events has been done by three survey groups, i.e., Massive Astrophysical Compact Halo Object (MACHO), EROS and OGLE, for more than one decade \citep{Alcock1997MACHOs,EROS1999,Udalski2000}. The aim of these observations was to search for MACHOs in the Galactic Dark Matter (DM) halo \citep[see, e.g., ][]{Alcock2000, Tisserand2007, Niikura2019}.

In this work, we aim to numerically study different aspects of detecting FFPs towards the Magellanic Clouds. We first do a Monte-Carlo simulation of short-duration microlensing events due to FFPs towards the MCs. We assume a log-uniform step function for FFPs' mass and study the efficiency of detecting FFPs by evaluating the signal-to-noise ratio (SNR) in several passbands and different surveys. We then promote our simulations to more realistic ones and assume that the \wfirst~telescope will continuously observe each MC during one observing $72$-day season with the $15$-min cadence. Our aim is to investigate the physical and statistical properties of detectable microlensing events.

The outline of the paper is as follows. In section \ref{two}, we briefly review some points for detecting FFPs through microlensing observations. In section \ref{three}, we explain our Monte Carlo simulation of microlensing events towards the Magellanic Clouds due to FFPs. We select the discernible events based on the SNR criterion and discuss their properties through different filters. We improve our simulations to be more realistic according to the \wfirst~strategy for detecting short-duration microlensing events and estimate the expected number of detectable FFPs towards MCs in section \ref{four}. In section \ref{five}, we summarize the conclusions.
%%%%%%%%%%%%%%%%%%%%%%%%%%%%%%%%%%%%%%%%%%%%%%%%%%%%%%%%%%%%%%%%%%%

\section{Microlensing events due to FFPs}\label{two}
Short-duration microlensing events are generated by either low-mass lenses (e.g., FFPs) or close ones or lenses with high transverse velocities. In all of these degenerate cases, the time scales of the resulting microlensing events are short. The microlensing time scale is given by \citep[see, e.g.,  ][]{schneider1992, Liping2005, Gaudi2012}:  
\begin{eqnarray}
t_{\rm E}=  \frac{R_{\rm E}}{v_{\rm{rel}}}=\frac{1}{v_{\rm {rel} } } \sqrt{\frac{4~G~M_{\rm l}~D_{\rm s}} { c^{2}} } \sqrt{x_{\rm{rel}}~(1-x_{\rm{rel}} )},
\end{eqnarray}
where, $R_{\rm E}$ is the Einstein radius which is the radius of the images ring for prefect alignment. $G$ is the gravitational constant, $c$ is the speed of light,  $x_{\rm{rel}}= D_{\rm l}/D_{\rm s}$ and $D_{\rm l},~D_{\rm s}$ are the lens and source distances from the observer, respectively. $v_{\rm{rel}}$ represents the size of the relative lens-source velocity, i.e., $v_{\rm{rel}}= |\bf{v}_{\rm{rel}}|$, where:  
\begin{eqnarray}\label{vrels}
\bf{v}_{\rm{rel}}= \bf{v}_{\rm l,~p} - \bf{v}_{\odot,~p} -\textit{x}_{\rm{rel}} \left(\bf{v}_{\rm s,~p}- \bf{v}_{\odot,~p}  \right),
\end{eqnarray}
\noindent where, the index $p$ refers to the component of velocities projected on the sky plane, $\bf{v}_{\odot}$ is the velocity vector of the Sun, $\bf{v}_{\rm l,~p}$ and $\bf{v}_{\rm s,~p}$  are velocity vectors of lens and source stars projected on the sky plane.

One of the major challenges in observations of short-duration microlensing events is the finite-source effect \citep[see, e.g., ][]{1994wittmoa,Nemiroff1994,Hamolli2015}. In short-duration microlensing events caused by FFPs, the Einstein radius is very small and comparable to the projected source radius on the lens plane, $R_{\rm E} \sim R_{\star} x_{\rm{rel}}$, where $R_{\star}$ is the source radius. The finite-source effect on magnification factor is evaluated by the normalized source radius, $\rho_{\star}$, which is given by:  
\begin{eqnarray}\label{rhos}
\rho_{\star}= \frac{R_{\star} x_{\rm{rel}}}{R_{\rm E}}=\frac{R_{\star} c}{\sqrt{4~G~M_{\rm l}~D_{\rm s}} } \sqrt{\frac{x_{\rm{rel}}}{1-x_{\rm{rel}}} }.
\end{eqnarray}
\noindent The magnification factor deviates from its simple form when $-1\lesssim \log_{10}[\rho_{\star}/u]\lesssim 2.0$ \citep[see, e.g., Fig. 2 in ][]{sajadian2020a}. Here, $u$ is the lens distance from the source centre projected onto the lens plane and normalized to the Einstein radius. If $\rho_{\star} \lesssim 0.1 u$, the magnification factor does not depend on $\rho_{\star}$. When $\rho_{\star} \gtrsim 100 u$, the finite-source size eliminates the lensing effect. 

\noindent For extremely large source sizes and when the lens transits the source surface, the magnification peak is estimated as \citep{1973Maeder, 1996GouldGaucherel, Agol2003}:

\begin{eqnarray}\label{finit}
A \simeq  1 + \frac{2}{\rho_{\star}^{2}}. 
\end{eqnarray}	

Generally, the finite-source effect in detecting and modeling microlensing events has some positive and negative aspects. Three of them are listed in the following.

\begin{itemize}
\item The finite-source effect in microlensing events reduces magnification peaks (see, Equation \ref{finit}). Consequently, the signal-to-noise ratio during microlensing events and, as a result, the probability for discerning them (specially when source stars are intrinsically faint) decreases.

\item The significant positive rule of the finite-source effect in microlensing events is related to breaking the microlensing degeneracy and measuring the angular Einstein radius $\theta_{\rm E}$. While modeling microlensing events the lens mass and its distance are evaluated if the angular Einstein radius and the relative lens-source parallax $\pi_{\rm{rel}}$ are measured. These two quantities are related to each other as:  
\begin{eqnarray}
\theta_{\rm E}= \sqrt{\kappa~M_{\rm l}~\pi_{\rm{rel}}}, 
\end{eqnarray}
\noindent where, $\kappa$ is a constant and $\pi_{\rm{rel}}=1~\rm{au}~(D_{\rm l}^{-1}- D_{\rm s}^{-1})$. The angular Einstein radius is:
\begin{eqnarray}
\theta_{\rm E}=\frac{R_{\rm E}}{D_{\rm l}}=\frac{\theta_{\star}}{\rho_{\star}}.
\end{eqnarray}
In current microlensing observations, $\theta_{\star}=R_{\star}/D_{\rm s}$ is estimated by probing de-reddened source positions upon the colour-magnitude diagram \citep{2004Yoo, 2011Bensby, 2013Nataf} and $\rho_{\star}$ is estimated by measuring the finite-source effect on microlensing lightcurves. Hence, the finite-source effect is required to measure $\theta_{\rm E}$ and potentially resolve the microlensing degeneracy.

\item Another rule of the finite-source size is related to the blending effect. This effect alters the brightness and angular radius of source stars as well as lensing parameters such as $t_{\rm E}$, $\rho_{\star}$, and $u_{0}$ (the lens impact parameter). As a result, the blending effect causes a degeneracy when modeling microlensing events. In microlensing events with strong finite-source effects, the blending effect will not change angular Einstein radii. Because it has the same effect on $\rho_{\star}$ as it does on $\theta_{\star}$. We note that the blending is still correlated with the Einstein crossing time and the lens impact parameter \citep{2020Mroz_1}.
\end{itemize}

The first (negative) issue of the finite-source effect restricts discovering short-duration microlensing events due to FFPs which is the subject of this paper. We note that two positive rules of the finite-source effect are related to resolving the microlensing degeneracy and indicating the lens mass, after discovering short-duration events \citep[see, e.g.,  ][]{2016Henderson}. 
	
The finite-lens effect is another (second-order) challenge in observing of short-duration microlensing events. This effect is highlighted when the lens size normalized to the Einstein radius is as large as the image positions projected on the lens plane and normalized to the Einstein radius \citep[see, e.g., ][]{Bromley1996, Bozza2002, Agol2002}. In single-lens microlensing events, the normalized image positions are:
\begin{eqnarray}
\theta_{\pm}= \frac{1}{2} (u \pm \sqrt{u^{2}+4}).
\end{eqnarray}
\noindent The size of the lens object normalized to the Einstein radius is:  
\begin{eqnarray}
\rho_{\rm l}= \frac{R_{\rm l}}{R_{\rm E}}.
\end{eqnarray}
\noindent The normalized lens size is too small in comparison to the image position ($\rho_{\rm l} \ll |\theta_{\pm}|$) for common microlensing events towards the Galactic bulge. But the finite-lens effect is considerable when either the lens object is large (e.g., red giants) or the Einstein radius is too small. In microlensing events towards the Magellanic Clouds due to low-mass microlenses the resulting Einstein radii are too small. Consequently, in our simulations we consider the finite-lens effect.  
%%%%%%%%%%%%%%%%%%%%%%%%%%%%%%%%%%%%%%%%%%%%%%%%%%%%%%%%%%%%%%%%%%%

\begin{figure}
\centering
\includegraphics[width=0.49\textwidth]{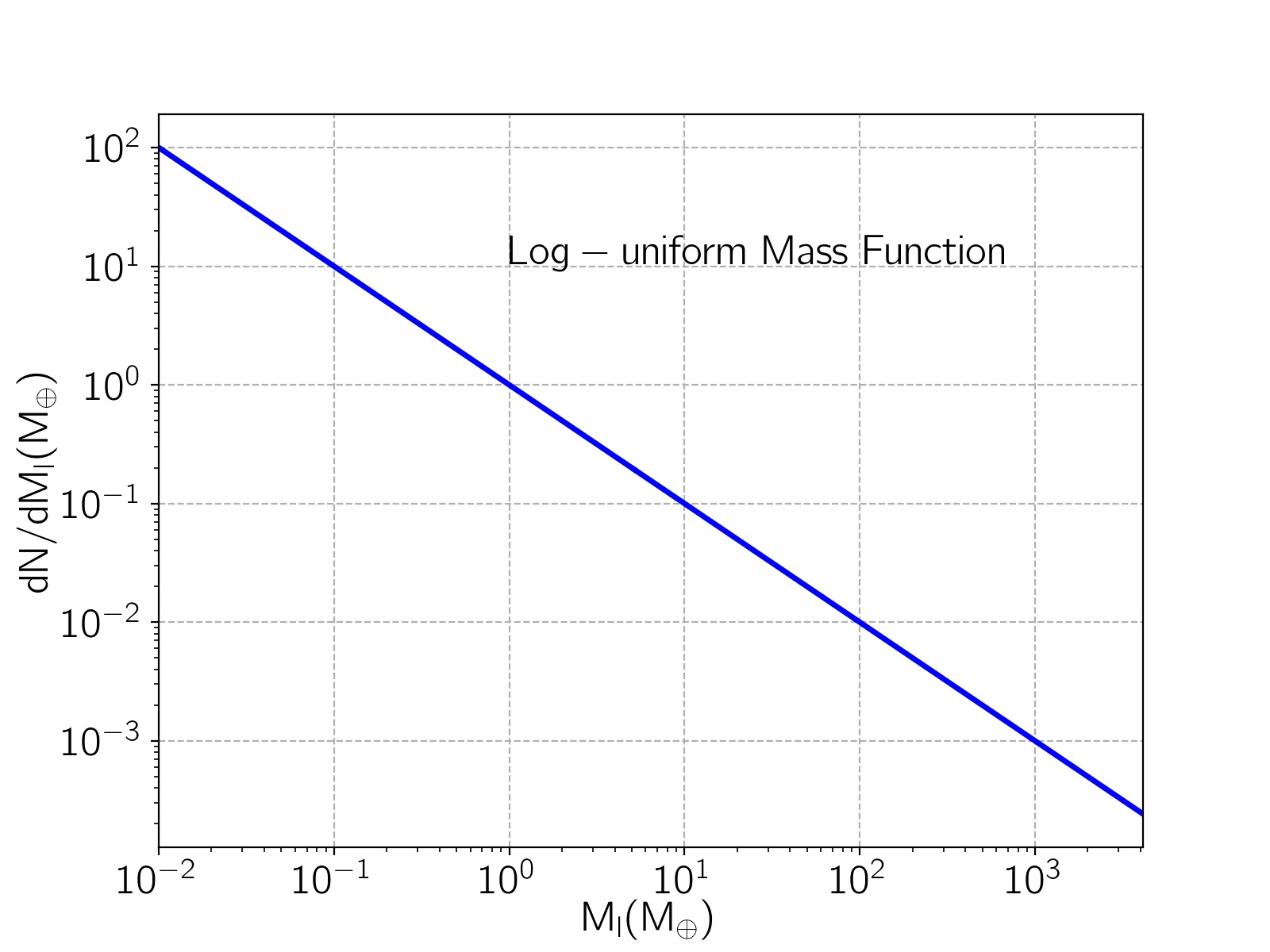}
\caption{The log-uniform mass function for FFPs used in our simulations.}\label{massfunc}
\end{figure}
\begin{figure*}
\centering
\includegraphics[width=0.49\textwidth]{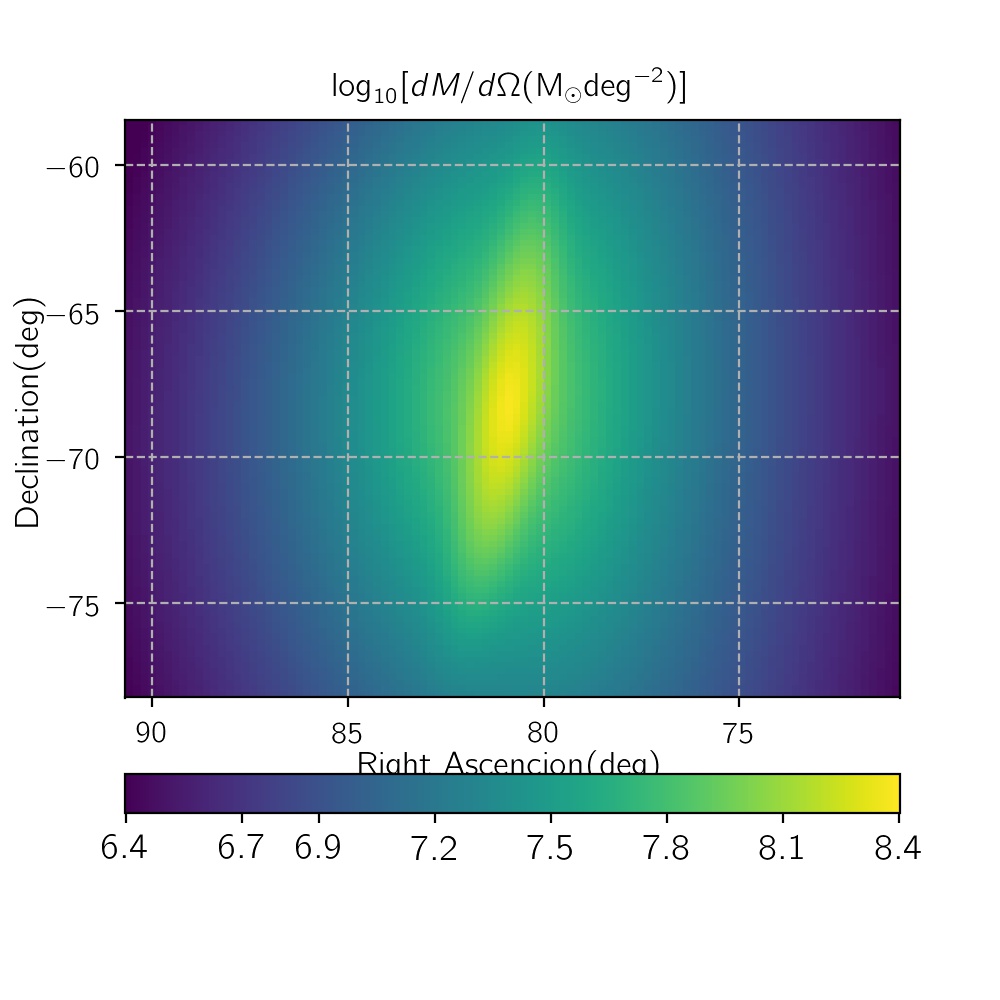}
\includegraphics[width=0.49\textwidth]{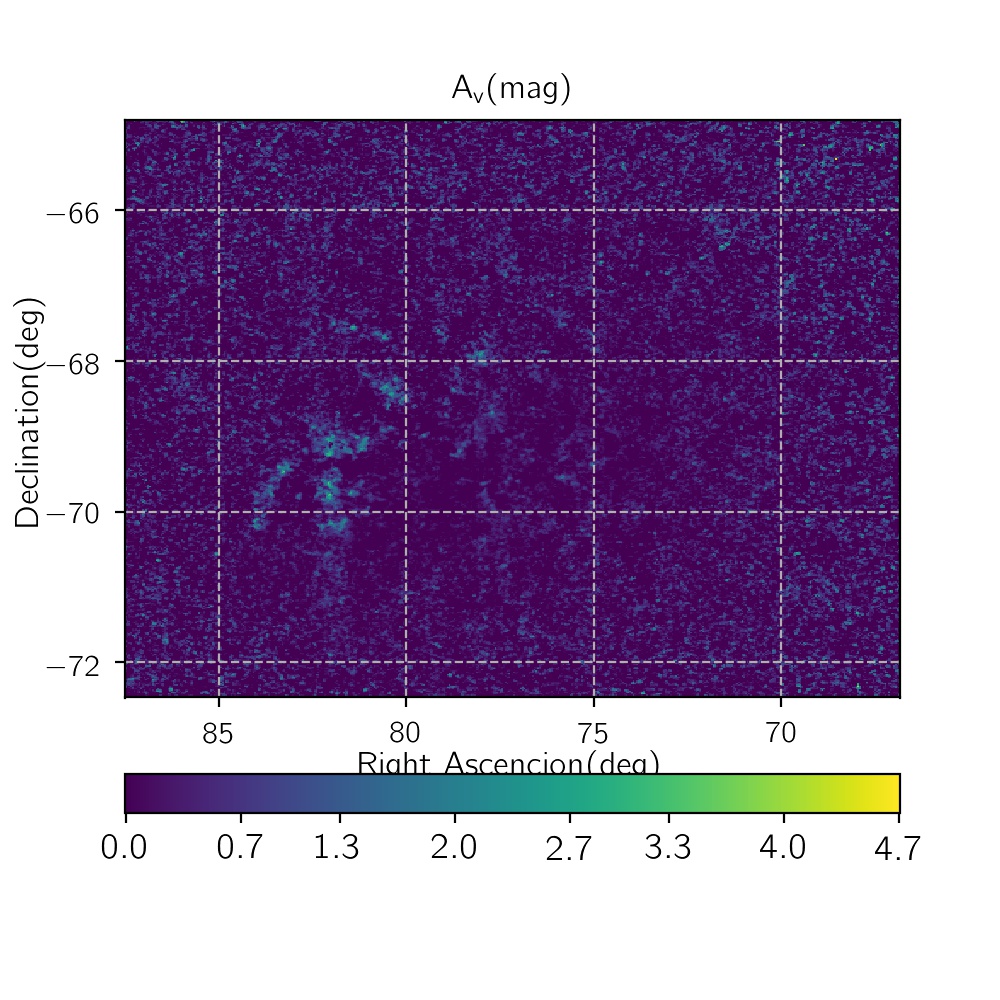}
\caption{Right panel: The map of the column mass density towards the LMC on the logarithmic scale in unit of $\rm{M_{\odot}deg^{-2}}$. Left panel: The map of the LMC $V$-band extinction, taken from \citet{ExtinctionLMC}. In these plots, the north and east directions are up and left, respectively.}\label{LMC_shape}
\end{figure*}
\section{Simulating Microlensing towards MCs}\label{three}
To perform a Monte-Carlo simulation of short-duration microlensing events towards the Large and Small MCs (LMC and SMC), we simulate a population of source stars and another population of lens objects and choose each source star and lens object for any microlensing event from these populations. Then, we determine lensing parameters. We aim to evaluate the properties of detectable short-duration microlensing events due to FFPs. In the following, we explain the physical distributions used to make lens and source star populations.  

The mass of lens objects is chosen from a log-uniform mass function, $dN/dM  \propto 1/M$, in the range $M_{\rm l}\in [0.01M_{\oplus},~13M_{\rm J}]$. This mass function is plotted in Figure \ref{massfunc}.   %We have used their known mass functions as \citep{SONYCII, Penny2019, Cassan2012}: 

The lens radius, $R_{\rm l}$, is a function of its mass \citep[see, e.g., ][]{massradius}. Recently, for a wide range of mass values and using a sample of $316$ well-constrained objects the mass-radius relation was well updated by \citet{2017Chen}, which is given in the following:  
\begin{eqnarray}
R_{\rm l} (R_{\oplus})= \nonumber~~~~~~~~~~~~~~~~~~~~~~~~~~~~~~~~~~~~~~~~~~~~~~~~~~~~~~~~~~~~~~~~~ \\
\begin{cases}\left[ M_{\rm l}(M_{\oplus})\right]^{0.28} &  M_{\rm l} < 2.0 M_{\oplus},\\
0.8 \left[ M_{\rm l}(M_{\oplus})\right]^{0.59}  &  2.0M_{\oplus} \leq M_{\rm l} < 0.4 M_{\rm J},\\
16.9 \left[ M_{\rm l}(M_{\oplus})\right]^{-0.04}  & 0.4M_{\rm J} \leq M_{\rm l} < 0.08M_{\odot},\\
0.0014\left[ M_{\rm l}(M_{\oplus})\right]^{0.88}  &  0.08M_{\odot} \leq M_{\rm l} < 2.0M_{\odot},\\
\end{cases}
\end{eqnarray}
where, the resulting radius is normalized to the Earth radius. This relationship is used in the simulation to determine the lens radius.

The lens object belongs to either our galaxy or the Magellanic clouds. The location of each lens is indicated by $d\Gamma/dx_{\rm{rel}} \propto \rho(x_{\rm{rel}}) \sqrt{x_{\rm{rel}}~(1-x_{\rm{rel}})}$,  where $\rho(x_{\rm{rel}})=\sum_{i=1}^{N} \rho_{i}(x_{\rm{rel}})$ is the cumulative mass density. Here, the summation is taken over mass densities in all structures of our galaxy and the MC (containing discs, bars and stellar halos). The mass densities in our galaxy are given by the Besan\c{c}on model\footnote{\url{https://model.obs-besancon.fr/}} \citep{robin2003,robin2012}. According to this model, the mass density in the Galactic disc has 8 components due to different stellar ages. The mass density in the Galactic bulge has two exponential components, as explained in \citet{robin2012}. The mass densities of the LMC and SMC stars are explained in Appendix \ref{append1}.  

In order to calculate the relative lens-source velocity, as given by Equation \ref{vrels}, we need the lens and source velocity vectors. The galactic stars have two types of velocities: global and dispersion velocities. The global rotation velocity is a function of the galacto-centric distance $r$, as \citep[e.g., ][]{2009Rahal}:
\begin{eqnarray}
v_{\rm{rot}}(r)=v_{\rm{rot},\odot} \left[ 1.00762 (\frac{r}{R_{\odot}})^{0.0394}  +  0.00712\right],
\end{eqnarray}	
where, $v_{\rm{rot}, \odot}=239\pm7\rm{ km~s^{-1}}$  and $R_{\odot}=8.0$kpc \citep{2011Brunthaler, Moniez2017}. The stellar dispersion velocities have three components, $v_{\rm R},~v_{\theta}, ~v_{\rm{Z}}$. We determine these components from Gaussian distributions. Their widths are related to the stellar age, the well-known age-velocity dispersion relation \citep[see, e.g., ][]{Jincheng2018, Mackereth2019}.  For the galactic stars, these widths are given in the Besan\c{c}on model. The lens impact parameter is chosen uniformly from the range $u_{0} \in [0,~1]$. The time of the closest approach is fixed at zero.

For the source stars, we first determine their distances from the observer using the overall mass density. In a given direction with specified right ascension and declination, $(\alpha,~\delta)$, we calculate the overall mass density versus distance, as
$$\frac{dM}{dD_{\rm s}}= D_{\rm s}^{2} \sum_{i=1}^{N} \rho_{i}(\alpha,~\delta,~D_{\rm s}) d\Omega.$$

\noindent In order to simulate the Magellanic Clouds and determine their mass densities in each given line of sight, we need two coordinate systems. (i) The observer coordinate system, $(x,~y,~z)$, with its centre on the Magellanic Clouds centre, $z$-axis pointing towards the observer, $x$-axis anti-parallel to the right ascension axis and $y$-axis parallel with the declination axis. (ii) The MC coordinate system, $(x',~y',~z')$, where its centre coincides with the Magellanic Clouds centre. $x'$ and $y'$ axes are parallel with the semi-major and semi-minor axes of the Magellanic Clouds' disc, respectively. Each point in the sky plane with the given equatorial coordinates (i.e., the distance $D$ from the observer and its right ascension and declination) is specified in the first coordinate system using the transformation relations in the spherical trigonometry \citep[see, e.g., ][]{Weinberg2001, VanderMarel2001a, Subramanian2012}. In the following subsections, \ref{lmc} and \ref{smc}, other details about simulating the Magellanic Clouds are explained.

\subsection{LMC}\label{lmc}

The major axis of the LMC's disc is at the position angle $\theta=170^{\circ}$ with respect to the $y$-axis. Therefore, we rotate the $x$-$y$ plane around $z$-axis by the angle $\theta-\pi/2$, so that the new $x'$-axis coincides with the major axis of the LMC \citep{Mancini2004}. Furthermore, the LMC disc has cylindrical symmetry, but it is inclined with respect to the sky plane by the angle $i=34.7^{\circ}$, giving the observer an elliptical shape. For converting to the Magellanic coordinate system, we then rotate the new system around its $x'$-axis by the inclination angle. The final conversion relation ( from the observer system to the MC one) is given in Appendix \ref{append1} and Equation \ref{convert}. In addition, the LMC has a bar at its centre. The position angle of the LMC bar is $\theta=110^{\circ}$ and with the zero inclination angle. For the stellar halos of the LMC and SMC, we assume mass densities with circular symmetry and no need to transform their coordinates. 

Here, for the LMC structures, we take the mass densities of the halo, disc and its bar from \citet{Kim2000}, as given in Appendix \ref{append1}. In addition, in the simulation we set the location of the LMC centre to:  
\begin{eqnarray}\label{lmcc}
(\alpha_{0},~\delta_{0})= (80.894^{\circ},~-68.244^{\circ}),\qquad  
D_{\rm{LMC}}=49.97~\rm{kpc}.
\end{eqnarray}

In the left panel of Figure \ref{LMC_shape}, the map of the LMC column density, in unit of $M_{\odot}~\rm{deg}^{-2}$, is shown, and is given by: 
\begin{eqnarray}\label{massd}
\frac{dM}{d\Omega}= \int_{D_{\rm s}} D_{\rm s}^{2} dD_{\rm s} \sum_{i=1}^{N} \rho_{i}(\alpha,~\delta,~D_{\rm s}).
\end{eqnarray}
\noindent The integration is carried over the distance $D_{\rm s}$ from the observer up to $65$ kpc. The resolution of this map is $0.2\times 0.2~\rm{deg}^{2}$. In this map, the north and east are towards up and left, respectively.

We assume the same (absolute) colour-magnitude diagrams for the source stars in the LMC as ones for the Galactic stars. The LMC's extinction map in the visible band is taken from \citet{ExtinctionLMC} and shown in the right panel of Figure \ref{LMC_shape}. The $V$-band extinction is converted to the extinction in other bands using \citet{Cardelli1989}'s relations. For the LMC stars the components of their global velocity are $V_{\rm R}=-57\pm 13$,~$V_{\rm T}=-226\pm 15$,~$V_{\rm Z}=221\pm 19$~$\rm{km}~\rm{s}^{-1}$ \citep{Kallivayalil2013,Moniez2020}.

\begin{figure}
\centering
\includegraphics[width=0.49\textwidth]{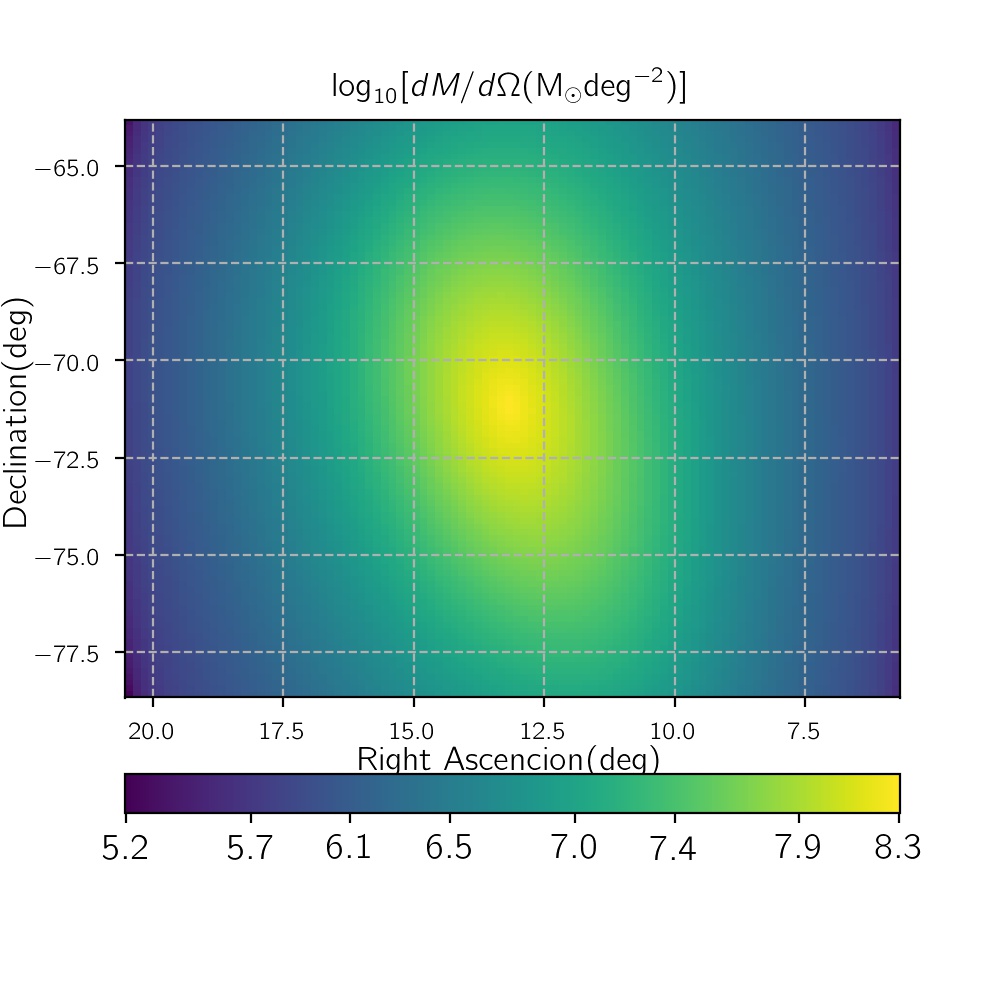}
\caption{The map of the column mass density towards the SMC in unit of $\rm{M_{\odot}deg^{-2}}$ and in the logarithmic scale. The resolution of the map is $0.15\times 0.15~\rm{deg}^{2}$.}\label{smcmass}
\end{figure}
\subsection{SMC}\label{smc}
Simulations of microlensing events in the SMC direction have been done in many references \citep[see, e.g., ][]{Graff1999, Wyrzykowski2011, Calchi2013, Mroz2018}. The SMC is at a larger distance in comparison to the LMC from the Sun. Its equatorial coordinates are: 
\begin{eqnarray}\label{smcc}
(\alpha_{0},~\delta_{0})= (13.187^{\circ},~-71.171^{\circ}),\qquad  
D_{\rm{SMC}}=61.70~\rm{kpc}. 
\end{eqnarray}
The SMC disc has two components, the young stellar population (YS) and the old stellar population (OS).  The projection angles for the first population are $\theta=114^{\circ}$ and $i=74^{\circ}$, while for the second they are $\theta=97^{\circ}$ and $i=0$ \citep{Calchi2013}. The details of the SMC mass densities are explained in Appendix \ref{append1}. The SMC does not have any bulge structure.  

In Figure \ref{smcmass}, the map of the SMC column density in the logarithmic scale is represented. For this plot, we integrate into the cumulative mass density from the observer location up to the distance $75$ kpc. The map has a resolution of $0.15\times 0.15~\rm{deg}^{2}$. 

The extinction map of the SMC was studied based on OGLE-III and OGLE-IV observations \citep{Haschke2011,Joshi2019} and \textit{Hubble Space Telescope} (\textit{HST}) observations \citep{HSTextSMC}. In this paper, we use the data of Table (3)  \footnote{\url{http://cdsarc.u-strasbg.fr/ftp/J/A+A/628/A51/table3.dat}} of \citet{Joshi2019}. We determine the $V$-band extinction from reddening factor using $A_{\rm v}=3.24 E(V-I)/1.4\rm{(mag)}$. The velocity components of the SMC stars and in the Galactic system are \citep{Kallivayalil2013}:
\begin{eqnarray}
V_{\rm R},~V_{\rm T},~V_{\rm Z}=+19\pm 18,~-153\pm 21,~153\pm 17 ~\rm{km~s^{-1}}.
\end{eqnarray}
%The velocities are in unit of  
\begin{table*}
\centering
\caption{The parameters to evaluate SNRs in different observing survey programs and filters (Equation \ref{SNR}).}\label{SNR_param}
\begin{tabular}{cccccc}\toprule[1.5pt]
& $V\rm{-Ground}$ & $I\rm{-Ground}$&	$K\rm{-Ground}$& $H\rm{-Space}$& $\rm{W149}\rm{-Space}$\\
\toprule[1.5pt]
$\rm{Potential}~\rm{Survey}$& $\rm{ROME/REA}$ & $\rm{OGLE}$ & $\rm{VISTA}$ & $Euclid$ &  $\wfirst$ \\
$\mu_{\rm{sky}} \rm{(mag. arcs^{-2})}$& $21.8$ & $19.7$ & $13.0$  & $21.5$ & $21.5$ \\
$\rm{FWHM} \rm{(arcs)}$& $0.39$ & $1.0$ & $1.6$ &  $0.4$ & $0.33$ \\
$\Omega_{\rm{PSF}} \rm{(arcs^2)}$& $0.12$ & $0.79$ & $2.01$ &  $0.13$ & $0.09$ \\
$m_{\rm{zp}}\rm{(mag)}$&$22.0$ & $20.4$ & $24.6$ & $24.9$ &  $27.6$\\ 
$t_{\rm{exp}}\rm{(s)}$& $200$ &  $180$ &  $60$ & $54$ & $46.8$ \\
$\delta_{\rm{m,~max}}\rm{(mag)}$& $0.09$ & $0.09$ & $3.67$ & $0.26$ & $0.60$\\ 
\hline
\end{tabular}
\end{table*}

%%%%%%%%%%%I%%%%%%%%%%%%%%%%%%%%%%%%%%%%%%%%%%%%%%%%%%%%%%%%%%%%%%%%%%%%%%
\begin{table*}
\centering
\caption{The average parameters of detectable short-duration microlensing events caused by FFPs towards the LMC with $\rm{SNR}>50$. Two rows of the physical parameters, for each filter, are due to the events with microlenses inside our galaxy (first, I) and the LMC (second, II).}\label{table1}
\begin{tabular}{cccccccccc}\toprule[1.5pt]
&$\left<D_{\rm l}\right>$&$\log_{10}[\left<\rm{M_{\rm l}}\right>]$&$\log_{10}[\left<R_{\rm E}\right>]$&$\left<v_{\rm{rel}}\right>$&$\left<t_{\rm E}\right>$&$\log_{10}[\left<\rho_{\star}\right>]$&$\log_{10}[\left<\rho_{\rm  l}\right>]$&$\log_{10}[\left<\rm{u_{0}}\right>]$&$\epsilon_{\rm{SNR}}[\%]$\\
&$(\rm{kpc})$&$(\rm{M_{\oplus}})$&$\rm{(au)}$&$\rm{(km/s)}$&$\rm{(days)}$&&&&\\	
\toprule[1.5pt]		
\multicolumn{10}{c}{$\rm{Ground-based}~ \textit{V}-{\rm{band}}$}\\
$\rm{I}$ & $0.794$ & $2.261$ & $-1.525$ & $47.720$ & $1.948$ & $-0.714$ & $-1.799$ & $-0.568$ & $0.686$\\
$\rm{II}$ & $48.774$ & $2.516$ & $-1.235$ & $40.384$ & $3.800$ & $1.400$ & $-1.919$ & $-0.416$ & $0.312$\\
\hline 
\multicolumn{10}{c}{$\rm{Ground-based}~ \textit{I}-{\rm{band}}$}\\
$\rm{I}$ & $0.770$ & $2.313$ & $-1.491$ & $48.212$ & $2.048$ & $-0.525$ & $-1.808$ & $-0.600$ & $0.294$\\
$\rm{II}$ & $48.692$ & $2.628$ & $-1.148$ & $40.241$ & $4.946$ & $1.778$ & $-1.943$ & $-0.388$ & $0.089$\\
\hline 
\multicolumn{10}{c}{$\rm{Ground-based}~ \textit{K}-{\rm{band}}$}\\
$\rm{I}$ & $0.772$ & $2.313$ & $-1.489$ & $48.350$ & $2.090$ & $-0.458$ & $-1.802$ & $-0.506$ & $0.317$\\
$\rm{II}$ & $48.743$ & $2.524$ & $-1.232$ & $40.662$ & $4.052$ & $1.758$ & $-1.895$ & $-0.335$ & $0.116$\\
\hline 
\multicolumn{10}{c}{$\rm{Space-based}~ \textit{H}-{\rm{band}}$}\\
$\rm{I}$ & $0.811$ & $2.241$ & $-1.537$ & $48.213$ & $1.879$ & $-1.244$ & $-1.793$ & $-0.549$ & $5.542$\\
$\rm{II}$ & $48.858$ & $2.427$ & $-1.310$ & $40.105$ & $3.108$ & $0.729$ & $-1.914$ & $-0.445$ & $3.528$\\
\hline 
\multicolumn{10}{c}{$\wfirst~ \textit{W149}-{\rm{band}}$}\\
$\rm{I}$ & $0.811$ & $2.241$ & $-1.537$ & $48.213$ & $1.879$ & $-1.244$ & $-1.793$ & $-0.549$ & $5.542$\\
$\rm{II}$ & $48.858$ & $2.427$ & $-1.310$ & $40.105$ & $3.108$ & $0.729$ & $-1.914$ & $-0.445$ & $3.529$\\
\hline 
\end{tabular}
\end{table*}
%%%%%%%%%%%I%%%%%%%%%%%%%%%%%%%%%%%%%%%%%%%%%%%%%%%%%%%%%%%%%%%%%%%%%%%%%%
\begin{figure*}
\centering
\includegraphics[width=0.49\textwidth]{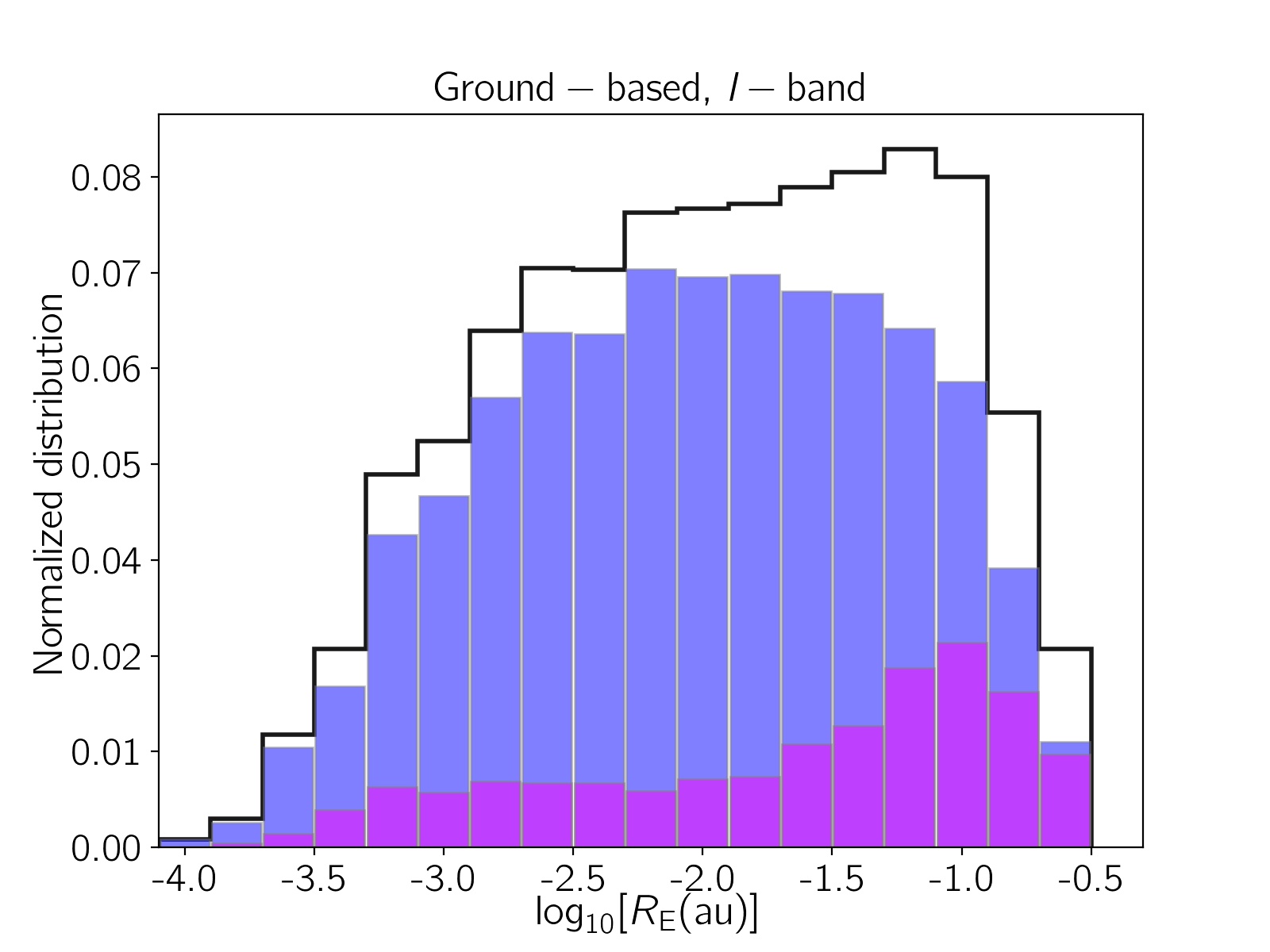}
\includegraphics[width=0.49\textwidth]{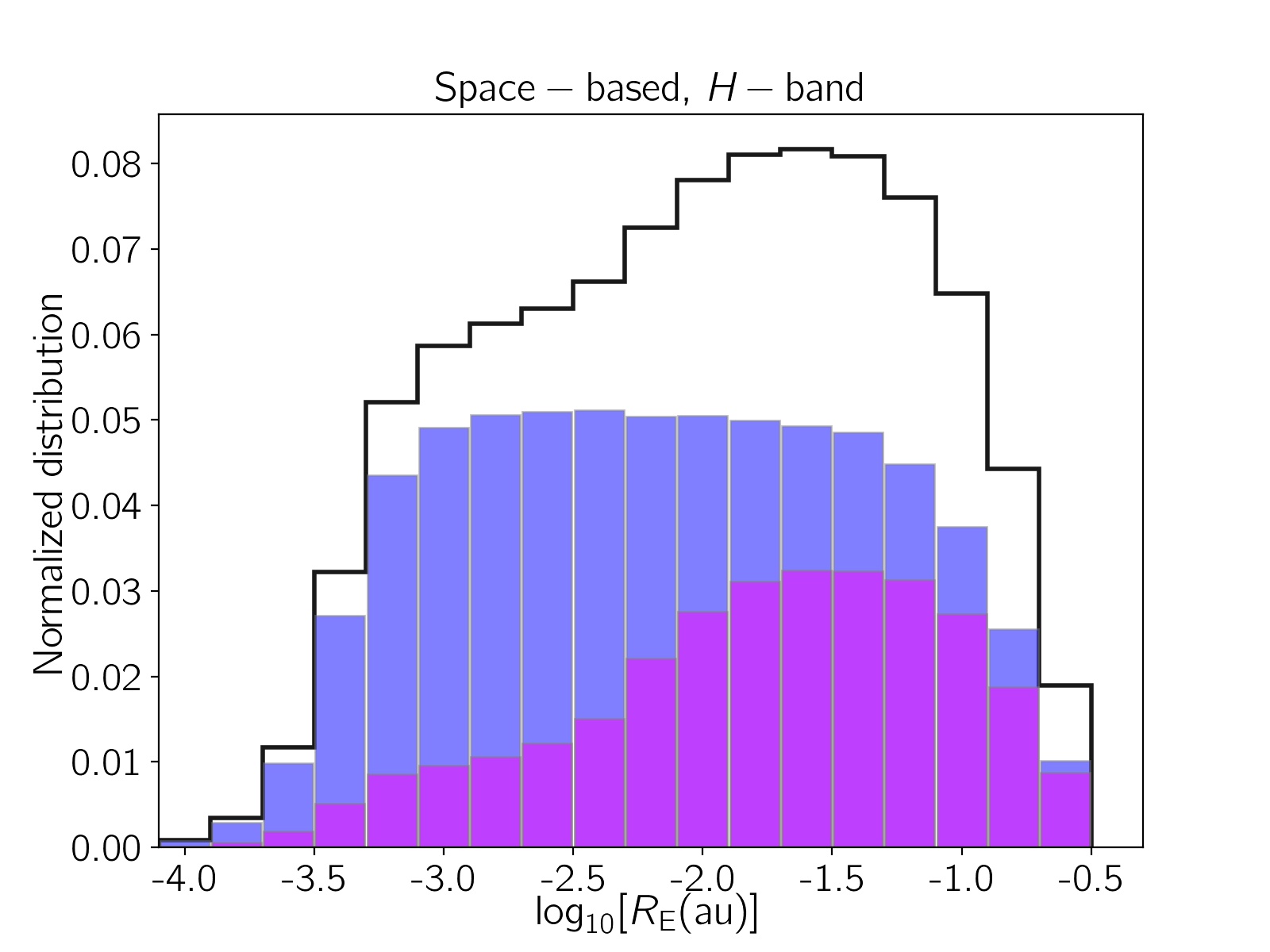}
\includegraphics[width=0.49\textwidth]{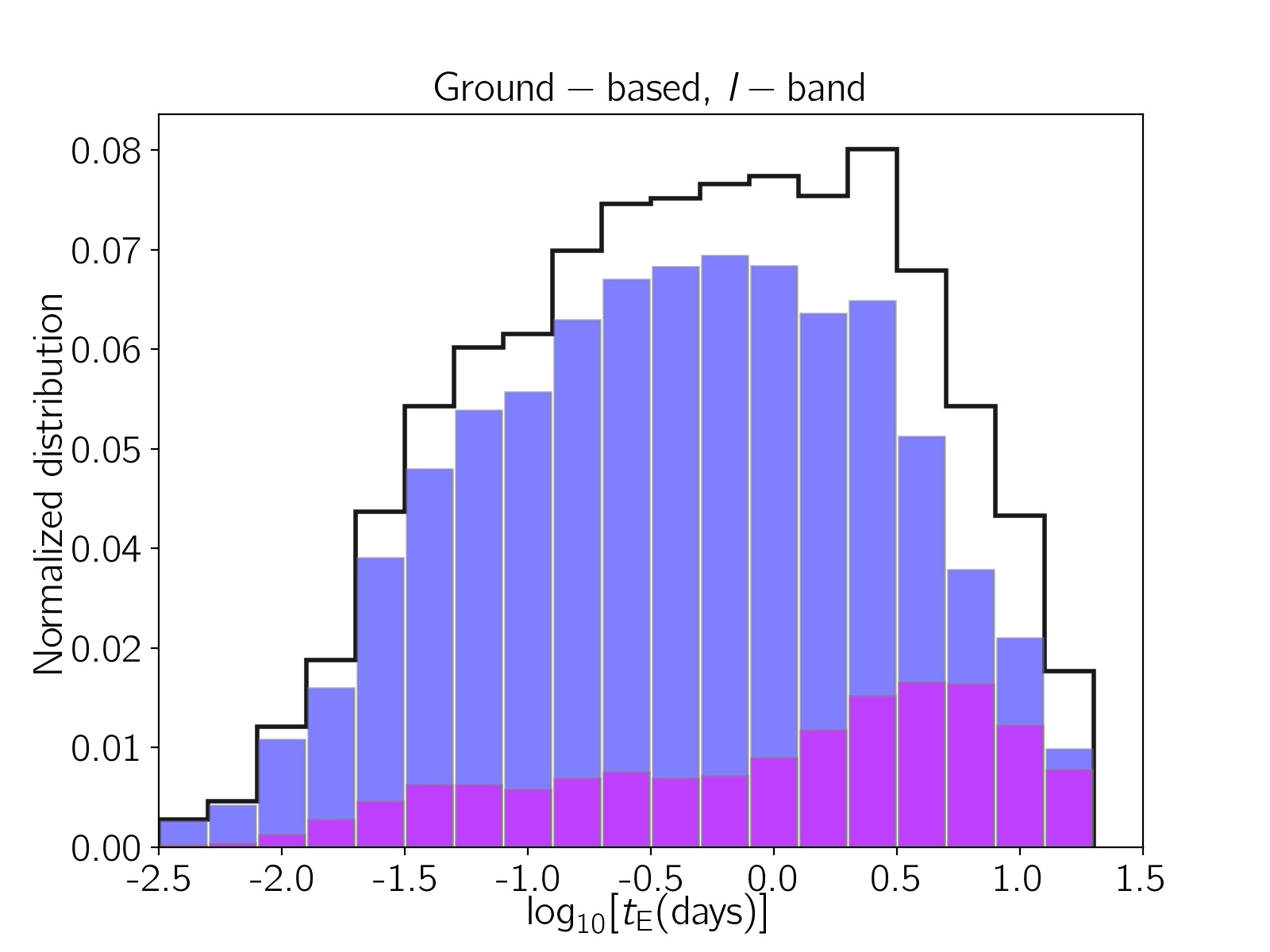}
\includegraphics[width=0.49\textwidth]{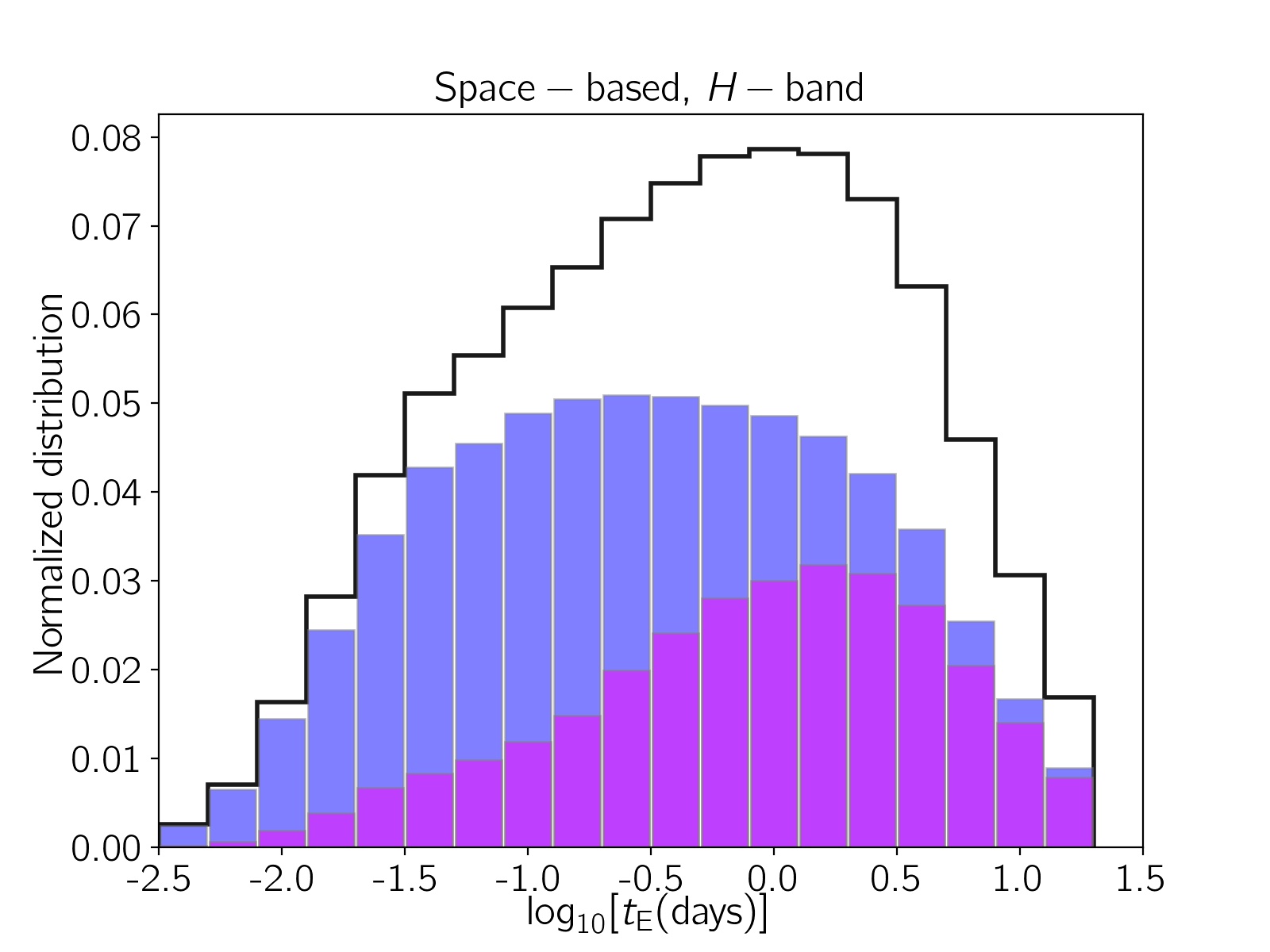}
\includegraphics[width=0.49\textwidth]{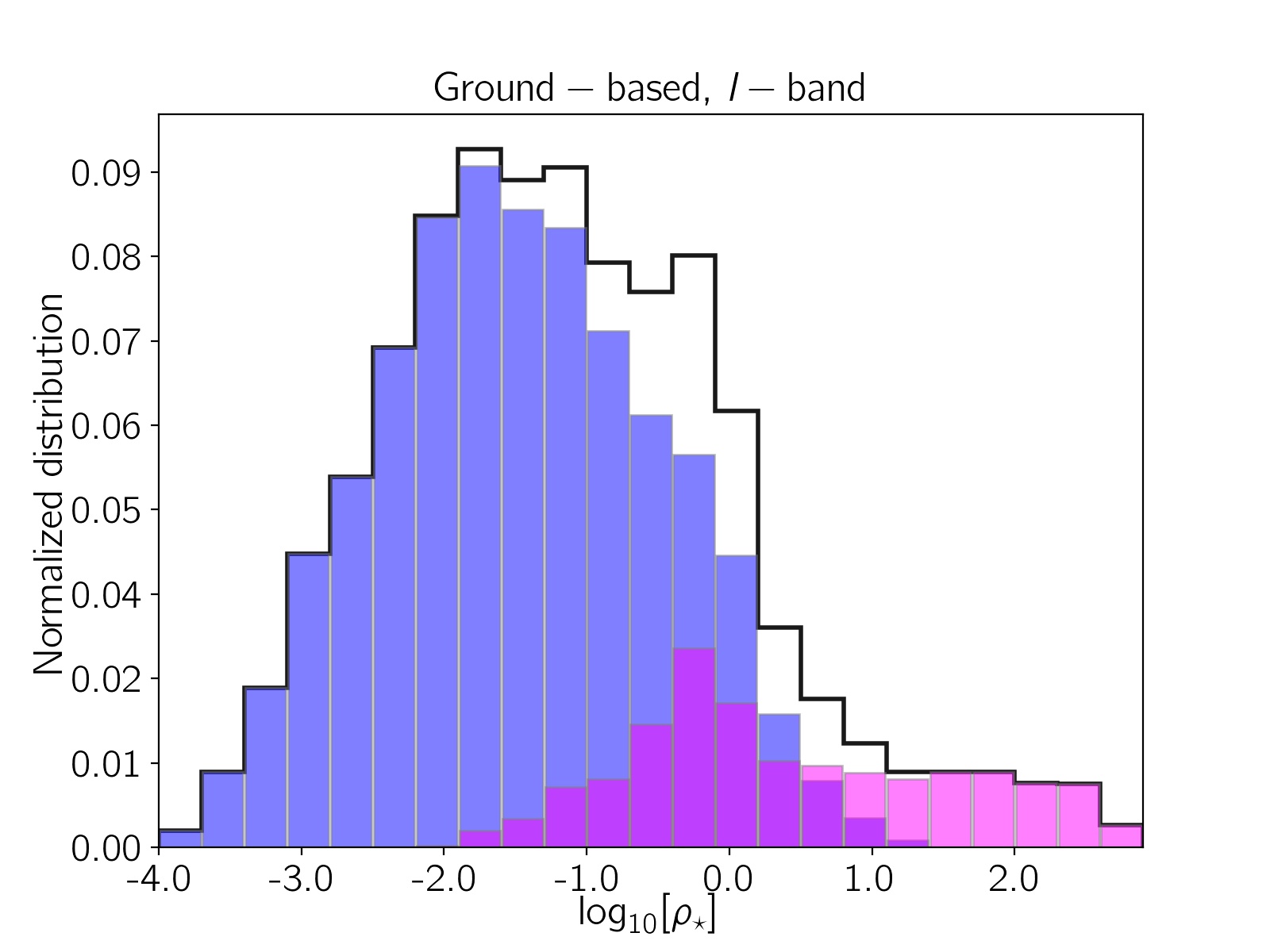}
\includegraphics[width=0.49\textwidth]{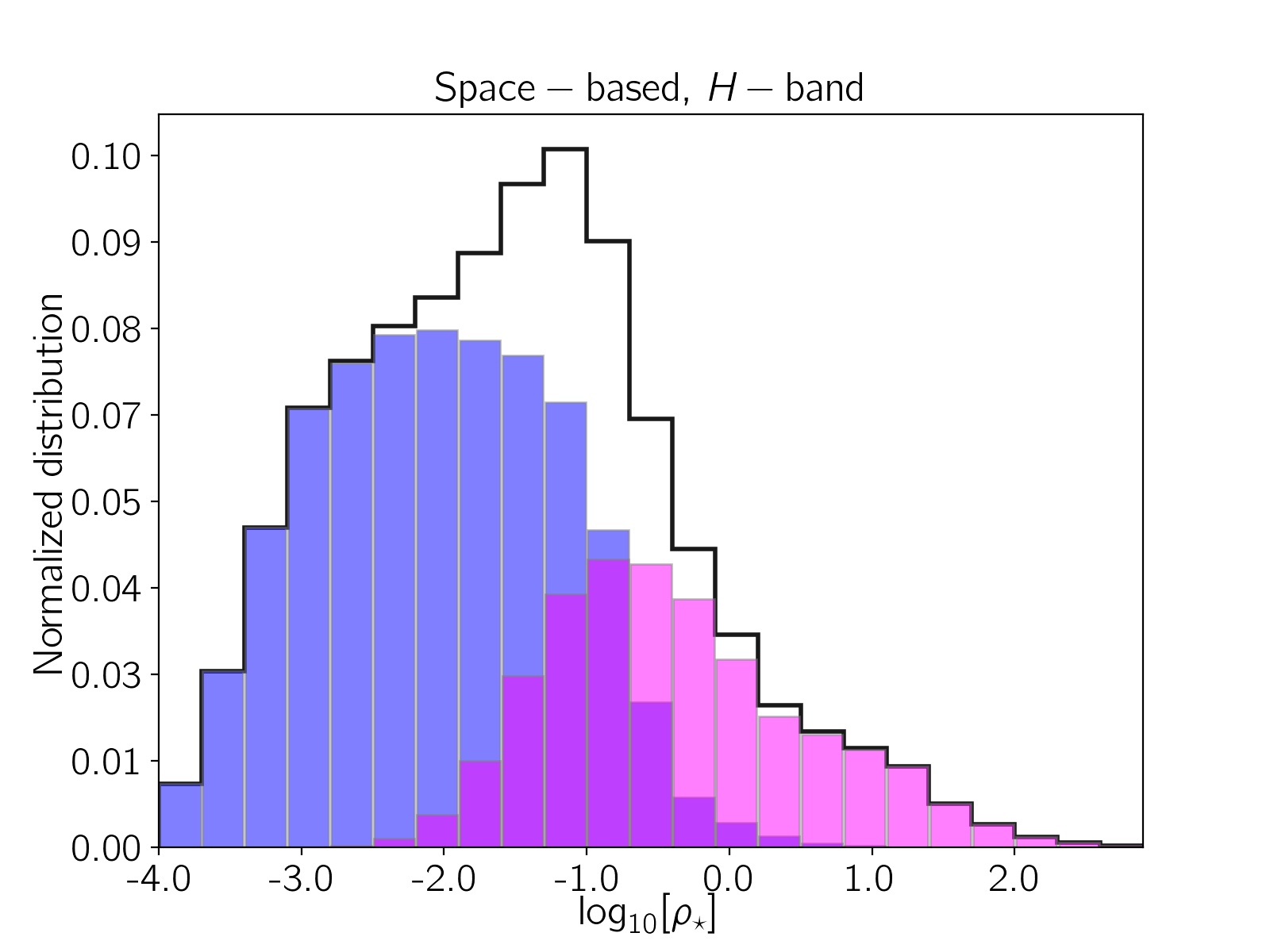}
\caption{From top tp bottom, the normalized distributions $R_{\rm E}$, $t_{\rm E}$ and $\rho_{\star}$ of detectable microlensing events I (blue), II (magenta) and total events (step black one). The histograms for two different survey observing programs, ground-based $I$-band (left) and space-based $H$-band (right) are plotted.}\label{histo}
\end{figure*}
%%%%%%%%%%%%%%%%%%%%%%%%%%%%%%%%%%%%%%%%%%%%%%%%%%%%%%
\subsection{SNR as detectability criterion}
In order to choose discernible events, we evaluate the Signal to Noise Ratio (SNR) in various passbands and survey observing programs, as \citep[see, e.g., ][]{sajadian2012, 2016sajadianlucky}:

\begin{eqnarray}\label{SNR}
\rm{SNR}= \frac{\sqrt{t_{\rm{exp}}}~A~10^{-0.2(2m_{\star}-m_{\rm{zp}})}}{\sqrt{10^{-0.4m_{\rm{b}}}+\Omega_{\rm{PSF}}10^{-0.4 \mu_{\rm{sky}}}+(A-1)10^{-0.4 m_{\star}} } },
\end{eqnarray}
\noindent where, $A$ is the magnification factor, $t_{\rm{exp}}$ is the exposure time, $m_{\star}$ is the apparent magnitude of the source star in each given waveband, $m_{\rm{zp}}$ is the zero point magnitude. $\mu_{\rm{sky}}$ is the sky brightness in $\rm{mag~arcs^{-2}}$, $\Omega_{\rm{PSF}}=\pi (\rm{FWHM}/2)^{2}$ and $\rm{FWHM}$ is the Full Width at Half Maximum of the stellar brightness profile.  

\noindent In Equation \ref{SNR}, $m_{\rm{b}}=-2.5 \log_{10}(\sum_{i=1}^{N_{\rm b}}F_{i})$ is apparent magnitude at the baseline which is due to all blending stars (including the source star itself) with fluxes $F_{i}$. $N_{\rm b}$ is the number of blending stars whose lights enter the source PSF (Point Spread Function). We calculate its average value from the column number density of stars as $\left<N_{\rm b}\right>=\Omega_{\rm{PSF}} \frac{dN}{d\Omega}$, and $dN/d\Omega$ is determined from the column mass density (Equation \ref{massd}) and by dividing the mass density of each structure to the average mass of its stars.  

Generally, parameters in Equation \ref{SNR} can be tuned according to characterizations of a given survey observing program. \citet{Ban2016} have differently adjusted the exposure time and the zero point magnitude to fix the highest achievable uncertainty $\delta_{\rm{m,~max}}$. They have estimated the event rate of short-duration microlensing events due to FFPs towards the Galactic bulge direction. They considered three wide filters, $I,~K$-bands for ground-based survey observations and $H$-band for space-based observations. Their tuned parameters in these wide filters have been reported in their Table (2). We take the same parameters of these wide filters to calculate the signal to noise ratios and rewrite them in Table \ref{SNR_param}.

Furthermore, we take into account ground-based survey observations in the $V$-band and space-based survey observations in the W149 filter. For observation in W149 by \wfirst,~ the zero point magnitude, PSF area, sky brightness, and the exposure time were given in \citet{Penny2019} and they are mentioned in the last column of Table \ref{SNR_param}.

\noindent An example of microlensing survey observation in the visible band is ROME/REA which is a global network of robotic telescopes. This network accesses to the LCO (Las Cumbres Observatory) southern ring of identical $1$m telescopes at Cerro Tololo Inter-American Observatory(CTIO), Chile, the South African Astronomical Observatory (SAAO), South Africa and Siding Spring Observatory (SSO), Australia \citep{ROMEREA, Street2019}. The visible sky brightness at CTIO is $21.8~\rm{mag~arcs^{-2}}$ \footnote{\url{http://www.ctio.noao.edu/noao/content/night-sky-background}}. The pixel size of $0.389$ arcs for $1$m telescopes in the LCO network results $\Omega_{\rm{PSF}}=1.1~\rm{arcs^{2}}$. Similarly to what \citet{Ban2016} proposed, we adjust the exposure time and $m_{\rm{zp}}$ to reach the same worst photometric precision as in the $I$-band (which occurs for the faintest detectable source star with the apparent magnitude $m_{\star}=m_{\rm{zp}}$), i.e., $\delta_{\rm{m,~max}}\simeq 0.09$ mag. Here, $\delta_{\rm{m,~max}}=-2.5 \log_{10}\left[1+ 1/\rm{SNR}_{\rm{min}}\right]$.The last row of Table \ref{SNR_param} contains $\delta_{\rm{m,~max}}$ for all observing strategies. For calculating them, we ignore the blending effect.

The stellar absolute magnitude in W149 can be expressed as $M_{\rm{W149}}=(K+H+J)/3$ which is the average of the stellar absolute magnitudes in $KHJ$ bands \citep{Montet2017}. We note that the Besan\c{c}on model provides the absolute magnitude of stars in the standard passbands $UBVIK$ but not magnitudes of stars in $HJ$ bands. The stellar absolute magnitudes in these bands are determined by their magnitudes in the $K_{\rm s}$-, $I$-bands, stellar age, and metalicity. We find linear relationships between $K_{\rm s}$-$H$ and $I$-$J$ absolute magnitudes for given stellar age and metalicty values using the Dartmouth isochrones \citep{2008Dotter}, which yield the absolute magnitudes in $H$-, $J$-bands. We set $K_{\rm s}=1.0526~K$.

In the simulation, we calculate the SNRs at the magnification peaks. As a result, $A$ is the magnification factor when $u=u_{0}$ and after accounting the finite-source and finite-lens effects. We calculate it using V.~Bozza's well-developed $\rm{RT}$-model \footnote{\url{http://www.fisica.unisa.it/gravitationastrophysics/RTModel.htm}} \citep{Bozza2018,Bozza2010,Skowron2012}. We calculate $\rho_{\rm l}$ to evaluate the finite-lens effect. If $\rho_{\rm l}$ be greater than $\sim 0.1$ the occultation by the lens object affects the lightcurves significantly. To evaluate this effect and occultation of the images' area, we compare $\rho_{\rm l}$ with $\theta_{\pm}(u \pm \rho_{\star})$ which represent the edges of the images. In all filters the threshold SNR value is $50$. 
\begin{table*}
\centering
\caption{Same as Table \ref{table1} but for the SMC.}\label{table2}
\begin{tabular}{cccccccccc} \toprule[1.5pt]
& $\left<D_{\rm l}\right>$ & $\log_{10}[\left<\rm{M_{\rm l}}\right>]$ & $\log_{10}[\left<R_{\rm E}\right>]$ & $\left<v_{\rm{rel}}\right>$ & $\left<t_{\rm E}\right>$ & $\log_{10}[\left<\rho_{\star}\right>]$ & $\log_{10}[\left<\rho_{\rm  l}\right>]$& $\log_{10}[\left<\rm{u_{0}}\right>]$ &$\epsilon_{\rm{SNR}}[\%]$\\
& $(\rm{kpc})$  & $(\rm{M_{\oplus}})$  & $\rm{(au)}$ & $\rm{(km/s)}$ & $\rm{(days)}$ & && & \\	
\toprule[1.5pt]
\multicolumn{10}{c}{$\rm{Ground-based}~ \textit{V}-{\rm{band}}$}\\
$\rm{I}$ & $1.090$ & $2.239$ & $-1.491$ & $144.582$ & $0.484$ & $-0.642$ & $-1.784$ & $-0.556$ & $0.597$\\
$\rm{II}$ & $59.395$ & $2.561$ & $-0.960$ & $51.688$ & $6.058$ & $0.983$ & $-2.231$ & $-0.455$ & $0.263$\\
\hline 
\multicolumn{10}{c}{$\rm{Ground-based}~ \textit{I}-{\rm{band}}$}\\
$\rm{I}$ & $0.994$ & $2.261$ & $-1.481$ & $141.420$ & $0.504$ & $-0.429$ & $-1.766$ & $-0.598$ & $0.249$\\
$\rm{II}$ & $59.298$ & $2.688$ & $-0.858$ & $52.465$ & $8.356$ & $1.376$ & $-2.259$ & $-0.467$ & $0.075$\\
\hline 
\multicolumn{10}{c}{$\rm{Ground-based}~ \textit{K}-{\rm{band}}$}\\
$\rm{I}$ & $0.977$ & $2.259$ & $-1.485$ & $140.300$ & $0.501$ & $-0.382$ & $-1.754$ & $-0.512$ & $0.271$\\
$\rm{II}$ & $59.326$ & $2.612$ & $-0.923$ & $52.180$ & $7.027$ & $1.382$ & $-2.222$ & $-0.388$ & $0.091$\\
\hline 
\multicolumn{10}{c}{$\rm{Space-based}~ \textit{H}-{\rm{band}}$}\\
$\rm{I}$ & $1.136$ & $2.238$ & $-1.493$ & $145.993$ & $0.477$ & $-1.147$ & $-1.789$ & $-0.553$ & $4.257$\\
$\rm{II}$ & $59.453$ & $2.403$ & $-1.089$ & $51.701$ & $4.335$ & $0.457$ & $-2.210$ & $-0.468$ & $2.732$\\
\hline 
\multicolumn{10}{c}{$\wfirst~\textit{W149}-{\rm{band}}$}\\
$\rm{I}$ & $1.136$ & $2.238$ & $-1.493$ & $145.993$ & $0.477$ & $-1.147$ & $-1.789$ & $-0.553$ & $4.257$\\
$\rm{II}$ & $59.453$ & $2.403$ & $-1.089$ & $51.701$ & $4.335$ & $0.457$ & $-2.210$ & $-0.468$ & $2.732$\\
\hline
\end{tabular}
\end{table*}
%%%%%%%%%%%I%%%%%%%%%%%%%%%%%%%%%%%%%%%%%%%%%%%%%%%%%%%%%%%%%%%%%%%%%%%%%%
\begin{figure*}
	\centering
	\includegraphics[width=0.49\textwidth]{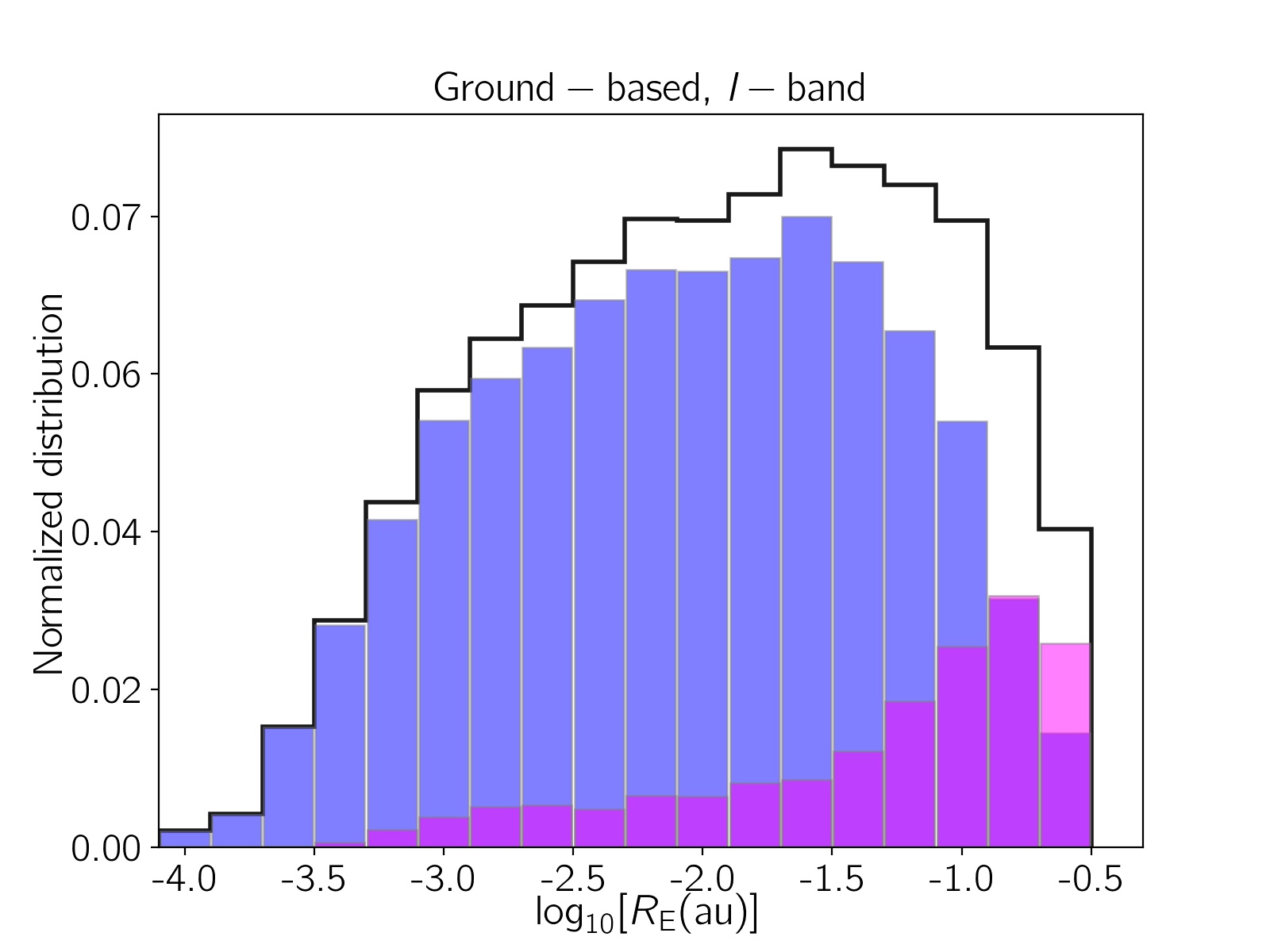}
	\includegraphics[width=0.49\textwidth]{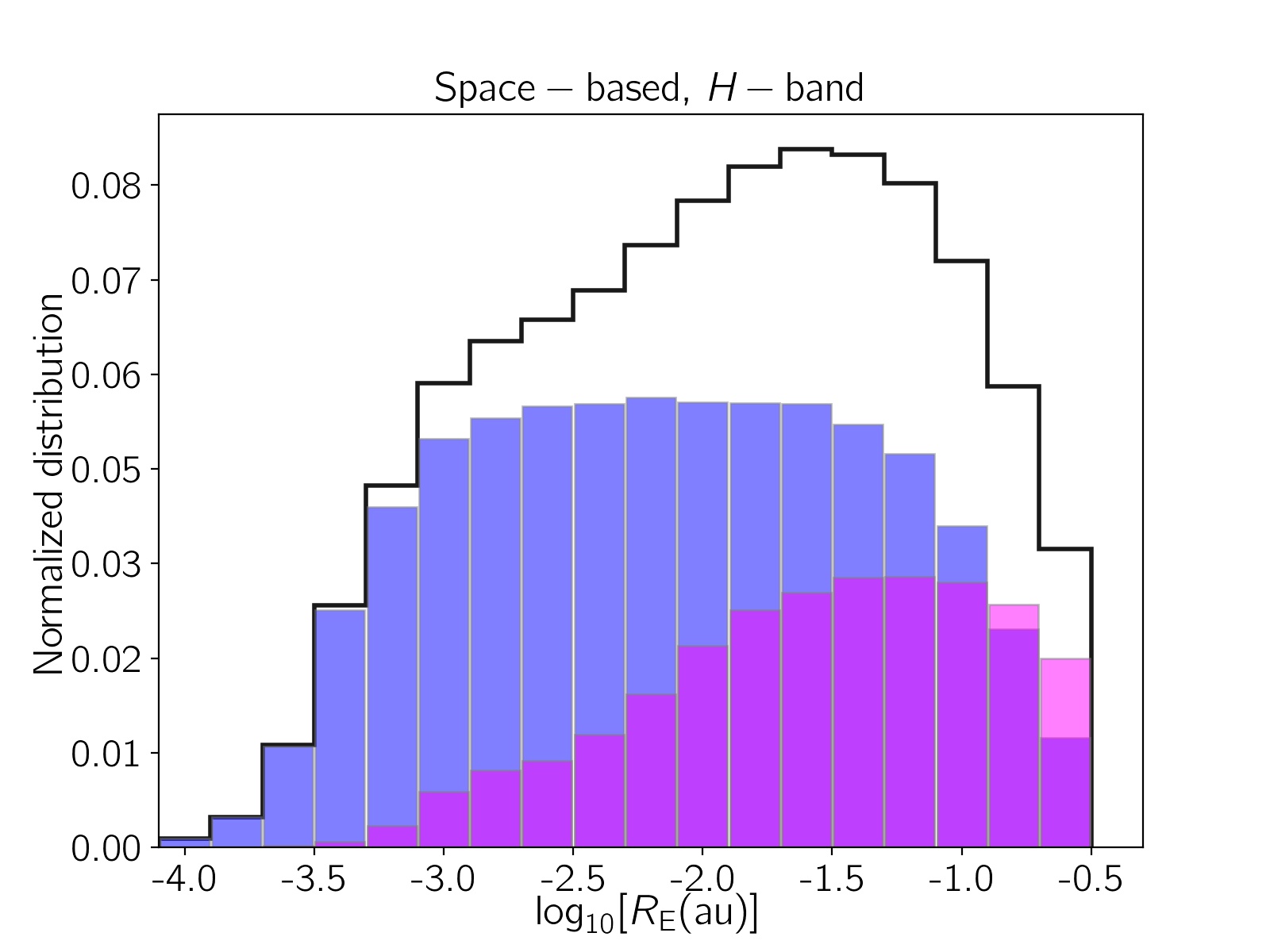}
	\includegraphics[width=0.49\textwidth]{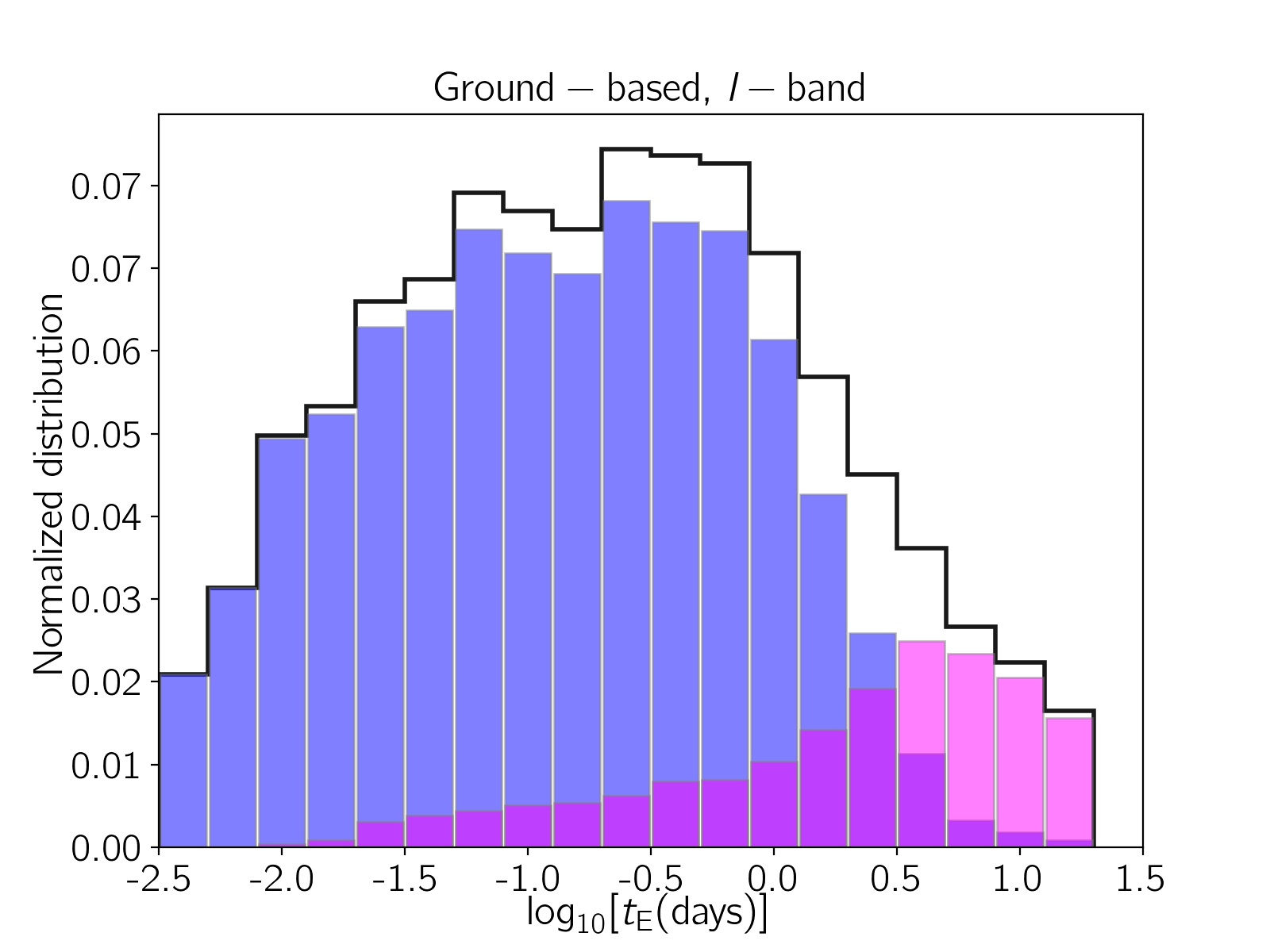}
	\includegraphics[width=0.49\textwidth]{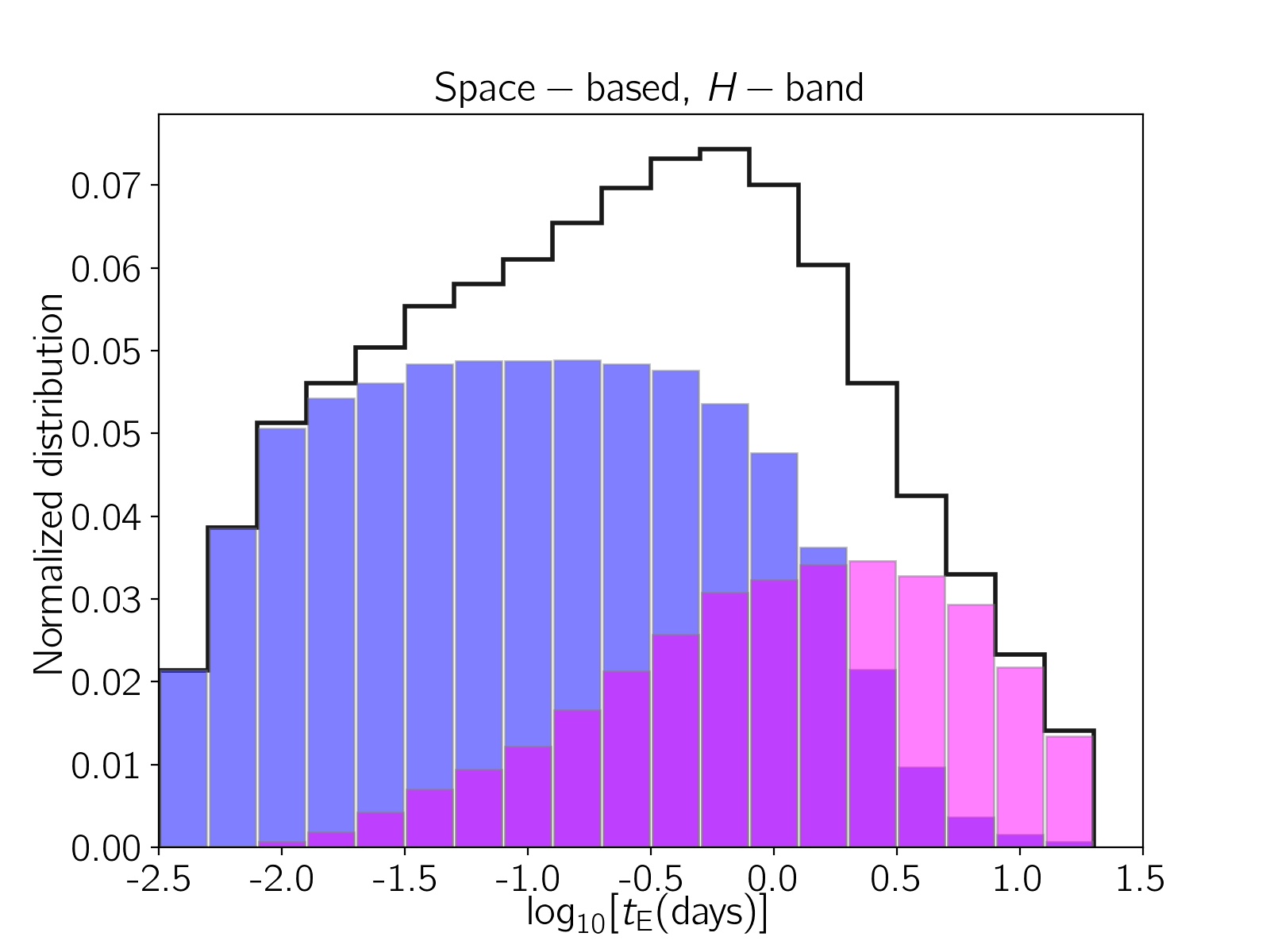}
	\includegraphics[width=0.49\textwidth]{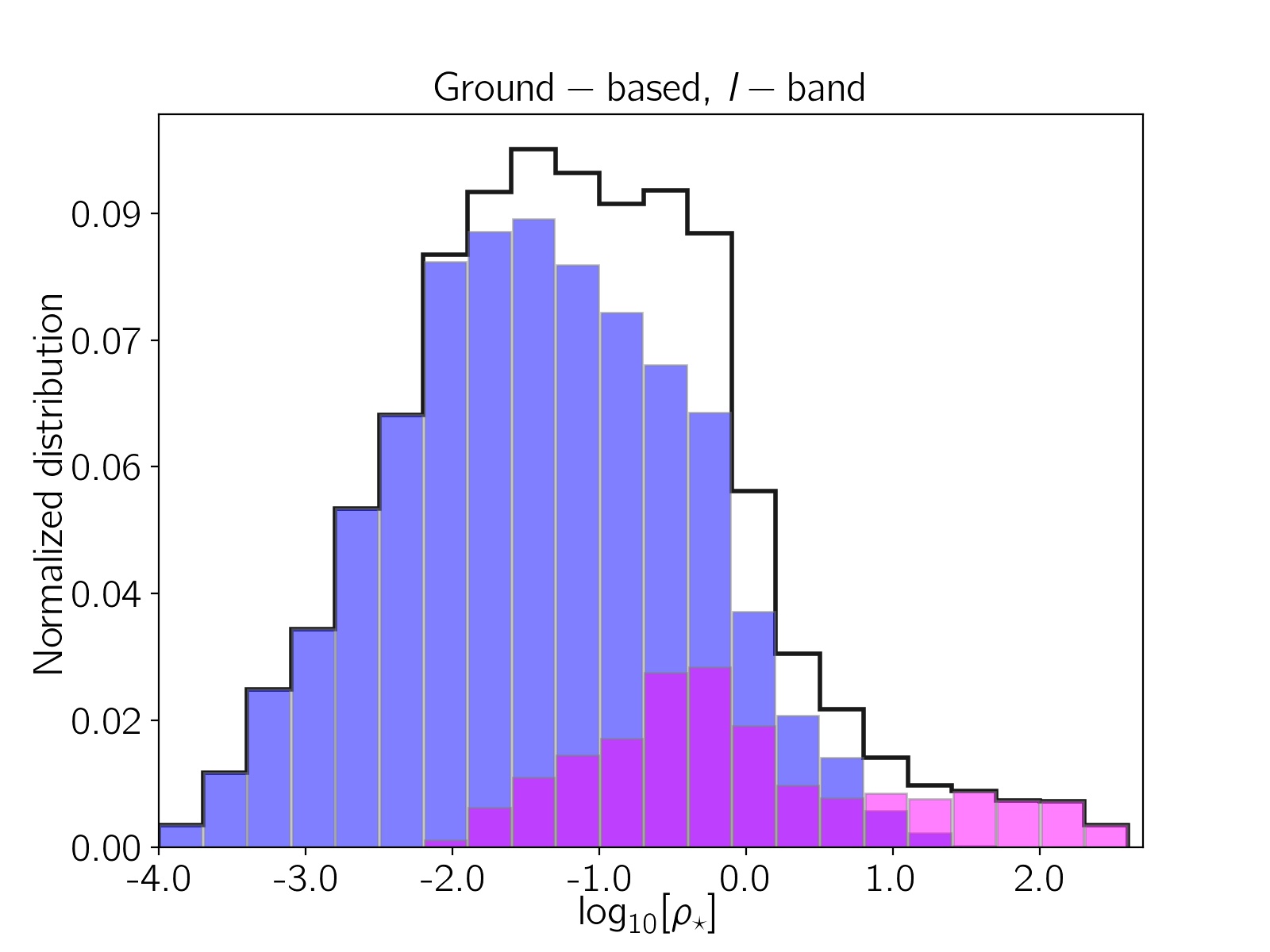}
	\includegraphics[width=0.49\textwidth]{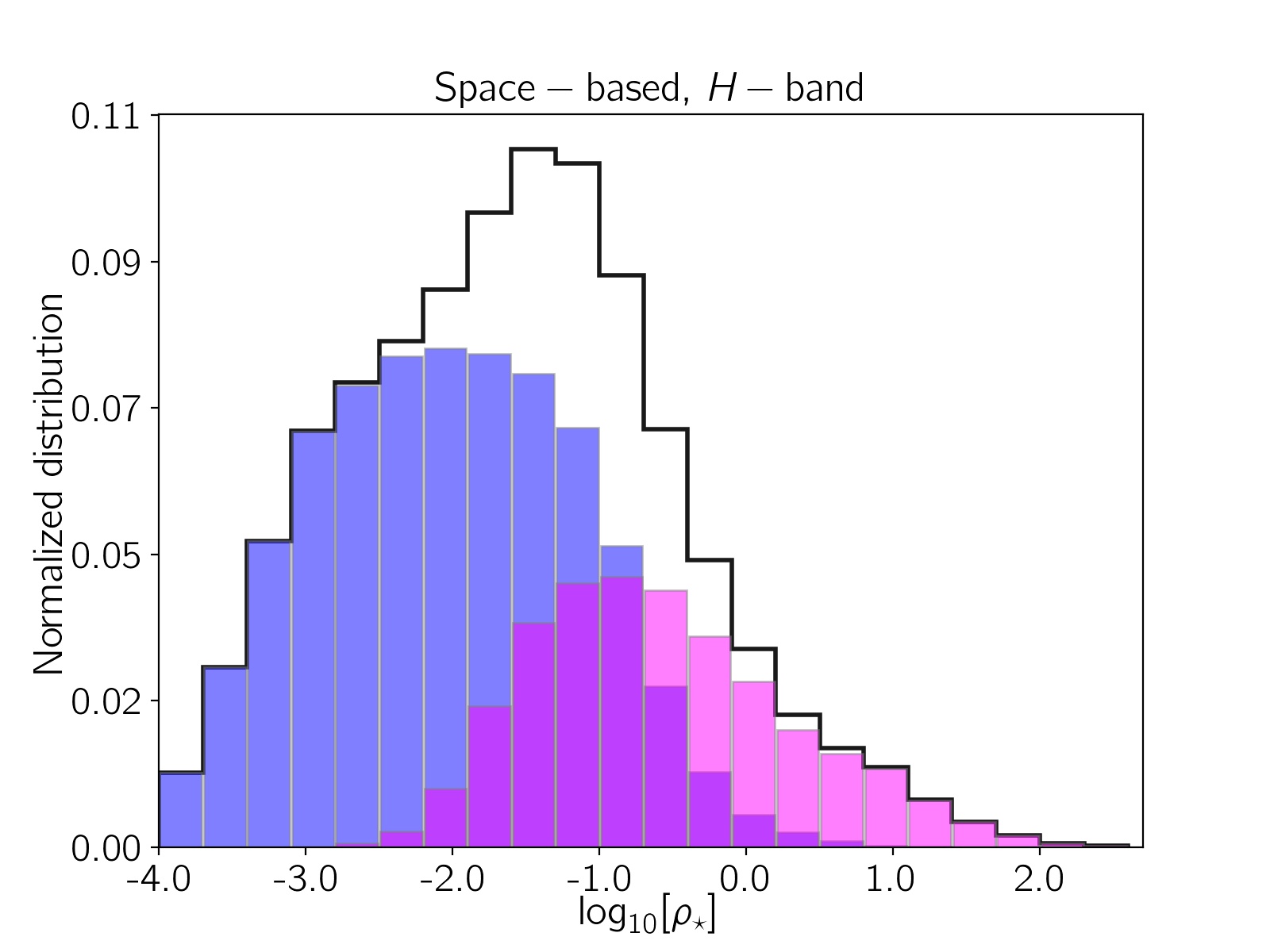}
	\caption{Same as Figure \ref{histo} but for the SMC.}\label{histo2}
\end{figure*}

\subsection{Properties of detectable events}
Following the Monte Carlo simulation, we separate the events with $\rm{SNR}>50$ (as detectable) and investigate their properties. 

\noindent The source distance in the Magellanic Clouds directions is $D_{\rm s} \gtrsim 50~$kpc, which is larger than the distance of the Galactic centre. For these observations, lens objects belong to either our galaxy or the Magellanic Clouds. For the first group, $x_{\rm{rel}}(=D_{\rm l}/D_{\rm s})\sim 0.02$-$0.07$ and for the second $x_{\rm{rel}} \sim 0.92$-$0.94$. The events with microlenses from our galaxy are referred to as  (I), while the events with microlenses from the LMC are referred to as (II). The ratio of their projected source radii for the same source star is $\rho_{\star, \rm{I}}/\rho_{\star, \rm{II}}\simeq 0.05$. Hence, these two samples of microlensing events (I,~II) have different finite-source effects. Additionally, because of different velocity distributions in our galaxy and the Magellanic Clouds, the time scales of events (I) and (II) are different.

\textbf{Results from simulating the LMC microlensing}: We represent the results of simulating microlensing events in the LMC direction in Table \ref{table1}. The table contains the average physical parameters of detectable microlensing events in different observing survey programs. We choose detectable microlensing events based on SNR calculations (Eq. \ref{SNR}). For each filter, two rows of parameters are given which are correspond to the detectable microlensing events (I) and (II), respectively. The last column in the table $\epsilon_{\rm{SNR}}$ denotes the fraction of detectable events (based on the SNR criterion) to total simulation events in each filter and lens location (I or II).  Some key points from this table are listed in the following. 

\begin{itemize}	
\item For a given lens mass, LMC events have similar $R_{\rm E}$ to those towards the Galactic bulge, but significantly lower relative velocities (less than half). Therefore, a given microlens in the Galactic halo generates longer LMC microlensing events with smaller finite-source effects in comparison to similar events towards the Galactic bulge. Although these two effects improve microlensing detectability, it additionally depends on stellar brightness and the blending effect. 
	
\item  An FFP with a given mass in our galaxy, (I), makes (on average) shorter (around half) events with smaller finite-source effects (by two orders of magnitude) than the events made by the same FFP in the LMC (II). To better show this point in Figure \ref{histo}, we plot the normalized distributions of the Einstein radius $R_{\rm E}$, the Einstein crossing time $t_{\rm E}$, and the projected source radius normalized to the Einstein radius, $\rho_{\star}$, for discernible microlensing events I (blue), II (magenta)  and total events (step black one) from top to bottom. These distributions are plotted for the detectable events in two survey programs, ground-based $I$-band (left) and space-based $H$-band (right). The peaks of these distributions due to events (I) and (II) have some displacements on the horizontal axis, which is more noticeable for $\rho_{\star}$ distributions. The finite-source effect decreases the efficiencies of detecting FFPs in events (II).     
	
\item Events (I) are detectable for smaller values of the lens impact parameters in comparison to events (II). Because events (I) have lower finite-source effects.  Based on the SNR criterion (which is related to the magnification peaks), on average, microlensing events (I) are more detectable than events (II).
	
\item We note that the smaller Einstein radius for events (I) offers higher finite-lens effects in comparison to events (II). The simulation shows that $\sim 0.1\%$ and $\sim 0.05\%$ of the simulated events whose microlenses belong to our galaxy and the LMC are affected by the occultation of source images through the finite-lens effect. However, this occultation effect does not significantly change detectability. 
\end{itemize}

%%%%%%%%%%%I%%%%%%%%%%%%%%%%%%%%%%%%%%%%%%%%%%%%%%%%%%%%%%%%%%%%%%%%%%%%%%
\textbf{Results from simulating the SMC microlensing}: In Table \ref{table2}, the results of performing the Monte-Carlo simulation of microlensing events due to FFPs in the SMC direction are represented. We use the SNR criterion to extract detectable events, i.e., $\rm{SNR}>50$, as explained in the previous subsection. Similarly, detectable microlensing events can be classified into two different groups. The events with microlenses inside our galaxy (I) and the events with  microlenses belong to the SMC (II). In the following, some key points are listed. 

\begin{itemize}
\item On average, the SMC events have relatively larger Einstein radii (because of the larger source distances) and higher relative lens-source velocities than the events towards the LMC. The effect of higher relative velocities is dominant for the events (I). Therefore, events (I) at the SMC are even shorter than those at the LMC. The effect of larger Einstein radii is dominant for the events (II) and these events towards the SMC are longer.

\item The SMC events (I) have larger finite-source effects than the LMC ones. The SMC events (II) have a smaller finite-source effect than the LMC ones.  

\item Although the SMC events (II) are longer and have smaller finite-source effects, the efficiency of detecting them is less than that towards the LMC. The SMC stars are on average fainter (due to larger distances) with higher blending effects which results lower SNR values and offers smaller efficiencies.

\item The relative lens-source velocity of events (I) is around $\left<v_{\rm{rel}}\right>\simeq 140~\rm{km s^{-1}}$, which is significantly higher than that due to the events (II). It causes that these events have a duration around $\left<t_{\rm E}\right>\simeq 12$ hrs, which is too short to be detected. On the other hand, the events (II) whose lens and source stars belong to the SMC are very rare (with a small contribution) because the cut radius of the SMC halo is $8.5$ kpc. Hence, the efficiency of detecting FFPs in the SMC observations is less than in the LMC ones.
\end{itemize}

We show the normalized distributions of the Einstein radius, the Einstein crossing time, and the normalized source radius from simulating microlensing events towards the SMC in Figure \ref{histo2}. The histograms for the events (I), (II), and total events are shown with blue and magenta colours and the step black function, respectively. These distributions have two peaks, as shown in Figure \ref{histo}. The SMC events with microlenses inside our galaxy are less affected by the finite-source effect, although their timescales are very short.

In general, via ground-based optical surveys (in the $VIK$ bands) the efficiencies to reach SNRs higher than $50$ are $\sim 0.3-0.6\%$ (for events with lenses inside our galaxy, I) and $\sim 0.1-0.3\%$ (for self-lensing events, II). These efficiencies are higher and are $\sim 4-5.5\%$ (I)  ($\sim 3-3.5\%$ (II)) when using space-based near-infrared surveys. Accordingly, the highest efficiency for capturing microlensing signals due to FFPs in the Galactic halo is achievable by a space-based near-infrared survey. The \wfirst~mission is a well-matched candidate for such observations and it is thoroughly investigated in the next section.

\noindent In the following section, we add synthetic data points (corresponding to \wfirst~observations of the MCs) to these simulated microlensing events for a more realistic selection of detectable events by \wfirst. We calculate the \wfirst~efficiencies for detecting the MCs microlensing events due to FFPs and estimate the number of observable events by using some other criteria (based on real observations) as detectability thresholds. 

\begin{figure*}
\centering
\includegraphics[width=0.49\textwidth]{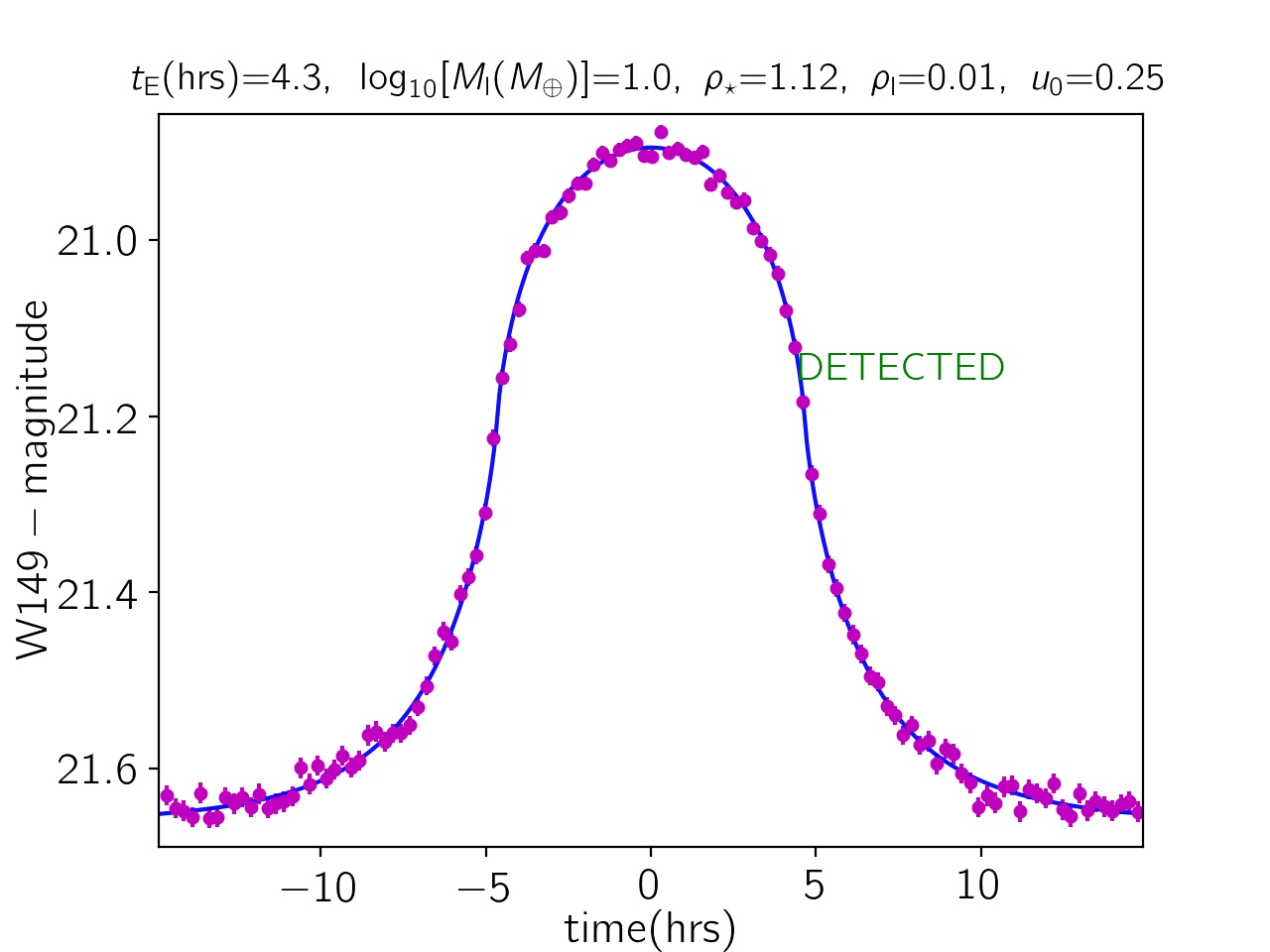}
\includegraphics[width=0.49\textwidth]{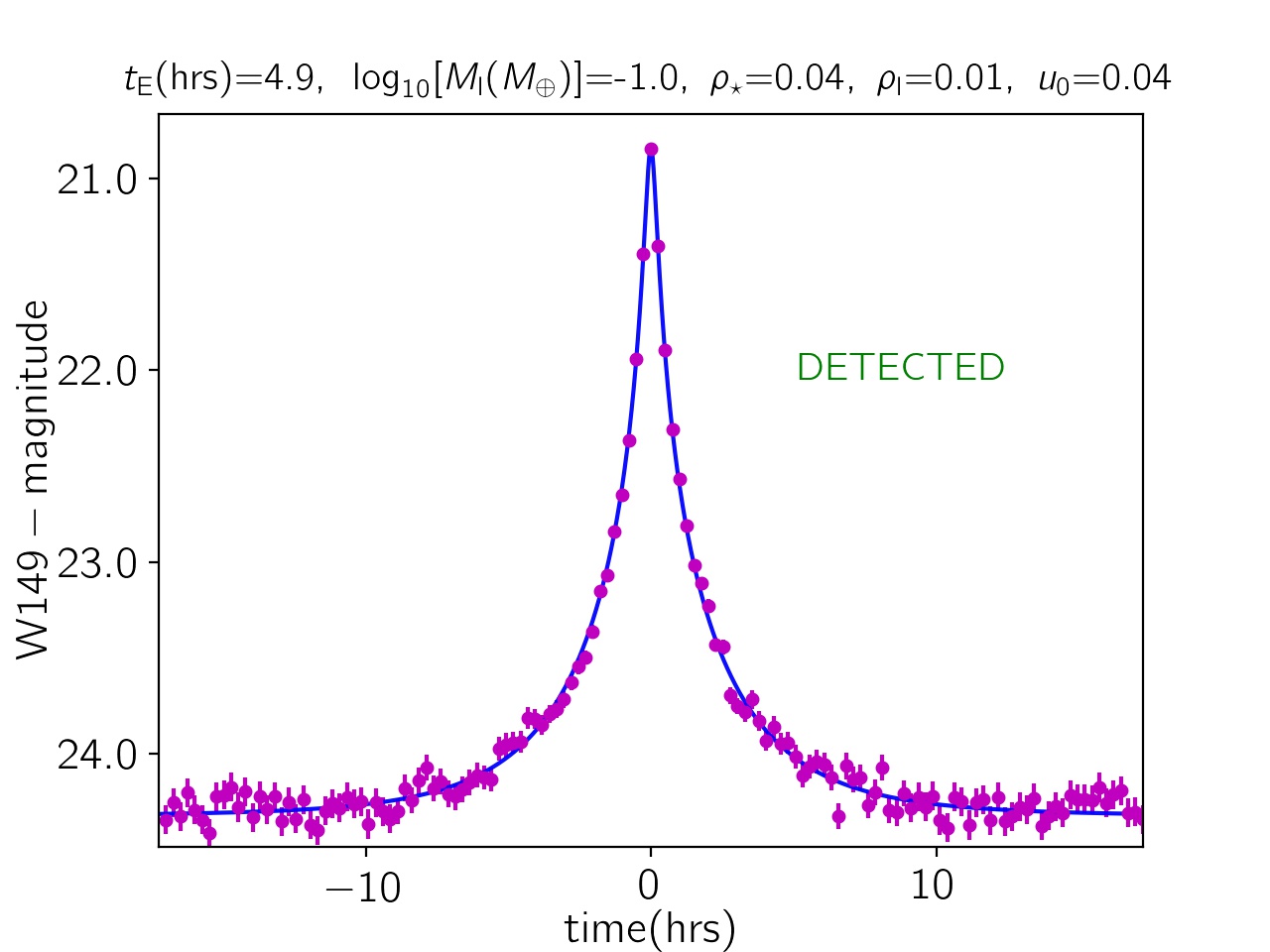}
\includegraphics[width=0.49\textwidth]{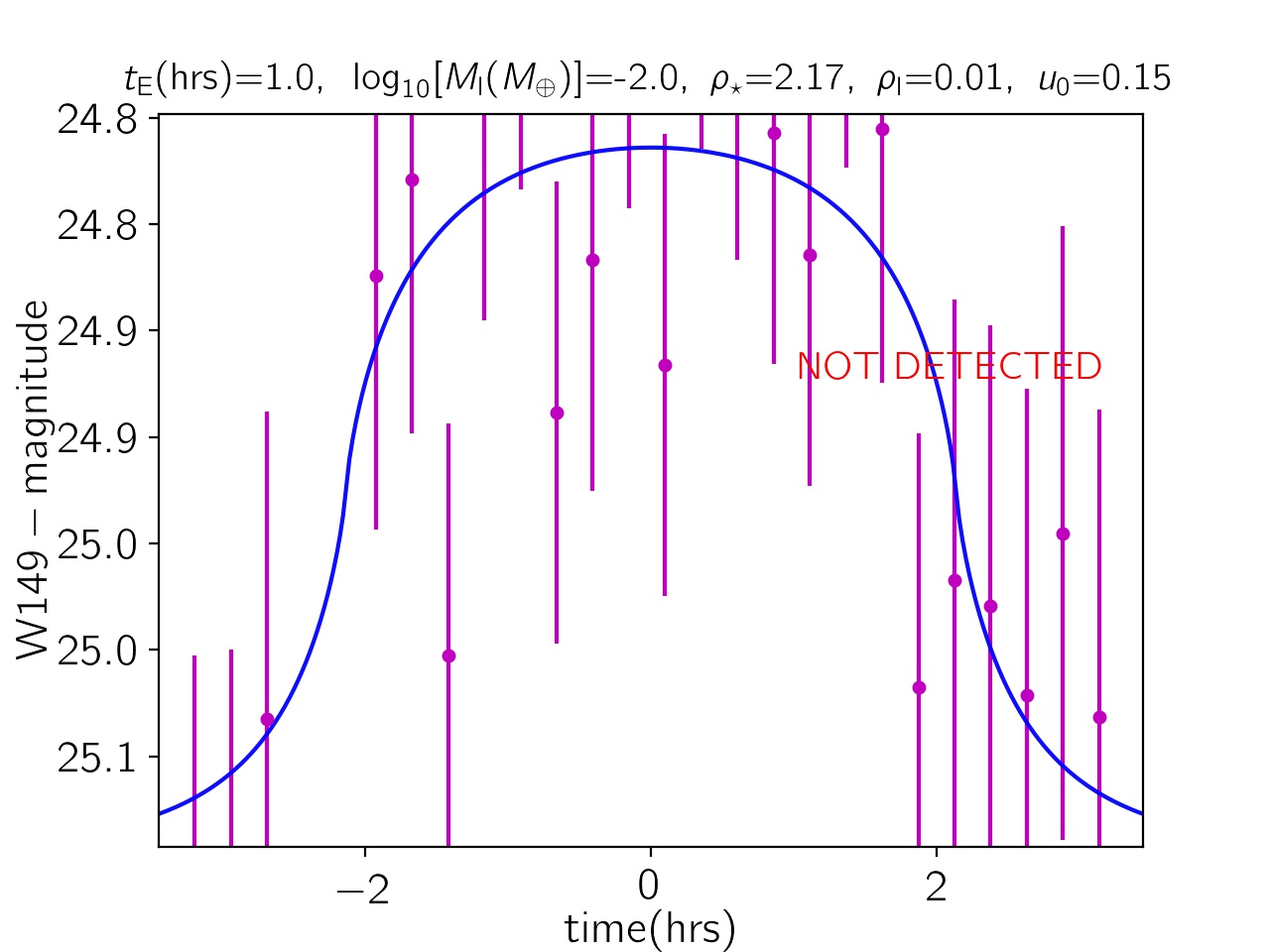}
\includegraphics[width=0.49\textwidth]{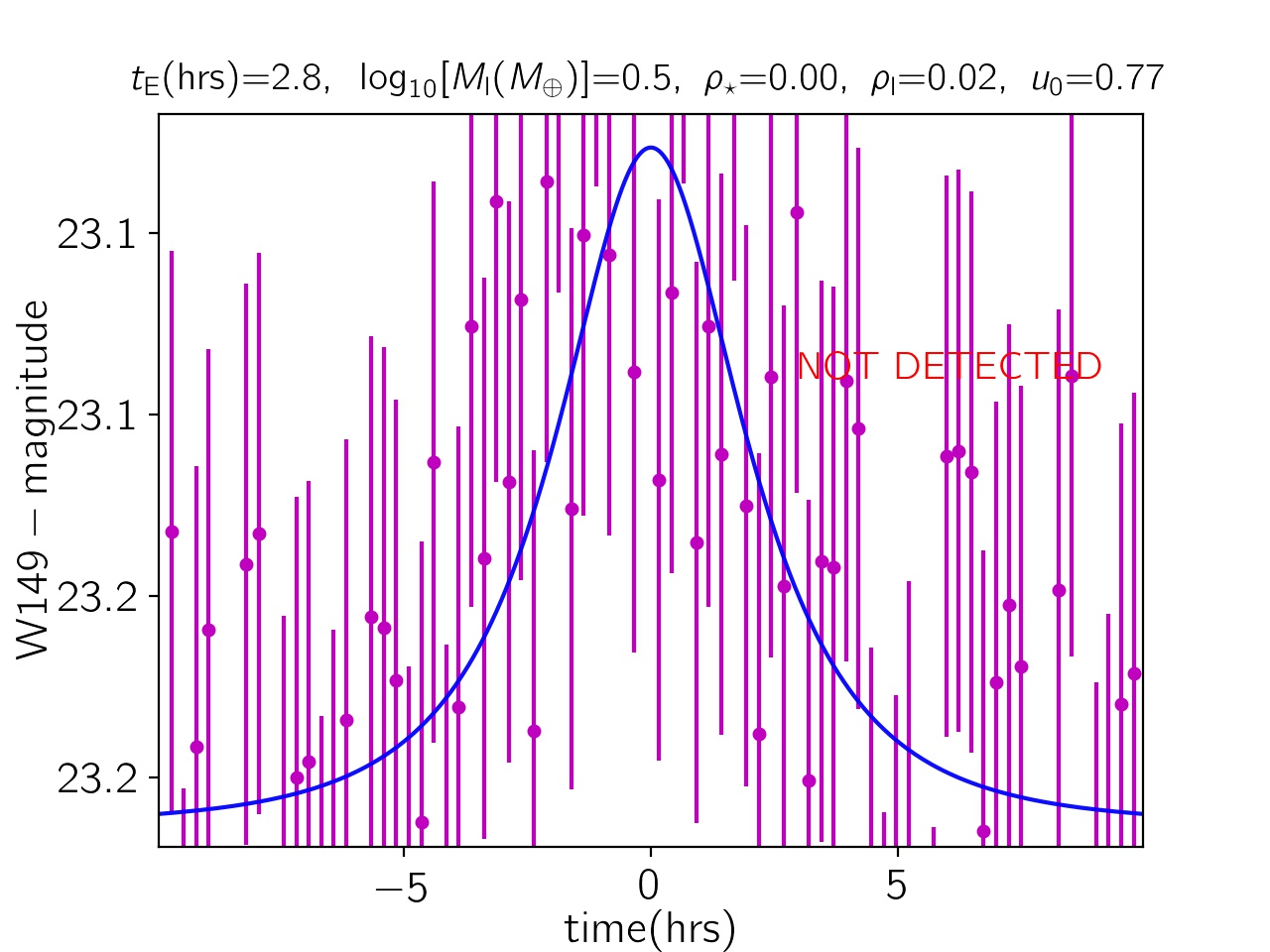}
\caption{Examples of the simulated microlensing lightcurves due to FFPs towards the Magellanic Clouds. The synthetic data points are represented by magenta-filled circles. The relevant parameters are mentioned at the top of each lightcurve.}\label{lightw}
\end{figure*}
%%%%%%%%%%%%%%%%%%%%%%%%%%%%%%%%%%%%%%%%%%%%%%%%%%%%%%%%
\section{\wfirst-based simulations}\label{four}
Here, we take a big ensemble of microlensing events that were simulated in the previous section, without applying any limitations. We assume that the \wfirst~telescope continuously observes these events during a $72$-day season and we generate synthetic data points that correspond to the \wfirst~observations. By applying some other criteria, we separate detectable events to indicate how many FFPs are potentially discovered through such observations.

\noindent To simulate observational data points, we assume a time interval of $15.16$ min between data points. The observation will be done with the W149 filter. For each data point, the apparent magnitude of the source star is indicated as:  
\begin{eqnarray}
m_{\rm{W149}}= m_{\rm{b},\rm{W149}} - 2.5 \log_{10} [f_{\rm b} A(u, \rho_{\star}, \rho_{\rm l}) +1 - f_{\rm b}],
\end{eqnarray}
where, $f_{\rm b}$ denotes the blending factor and $m_{\rm{b},\rm{W149}}$ is the baseline apparent magnitude in W149. The data points with the apparent magnitudes between the detection threshold and saturation limit, i.e., $m_{\rm{W149}}\in[14.8,~26]$ mag, are recorded. For each data point, the average uncertainty in the apparent magnitude, i.e., $\sigma_{\rm m}$, is indicated according to Figure (4) of \citet{Penny2019}. We uniformly select the time of the closest approach, $t_{0}$, from the range $[0,~72]$ days. The data points are simulated over a time interval of $[-3.5 t_{\rm E}+t_{0}, 3.5 t_{\rm E}+t_{0}]$.

In order to select discernible microlensing events, we consider three criteria. (i) the observing data should cover the magnification peaks, (ii) at least six data points should be higher than the baseline by more than $3 \sigma_{\rm m}$ and (iii) the difference between $\chi^{2}$ values from fitting the real model and the baseline without any lensing effect should be greater than a threshold amount, i.e., $\Delta \chi^{2}=|\chi^{2}_{\rm{real}}-\chi^{2}_{\rm{base}}| >300$. The events that pass all of these criteria are discernible in the \wfirst~observations. 

In Figure \ref{lightw}, four examples of the simulated lightcurves potentially observed by \wfirst~are shown. The magenta circles represent the synthetic data points collected by \wfirst.~The parameters of lightcurves are mentioned at the top of them. Two first lightcurves are discernible in the \wfirst~observations, although the first one is affected by the finite-source effect. Two other lightcurves are not confirmed by \wfirst.

The event rate and the expected number of events are calculated using the following relationships, respectively: 
\begin{eqnarray}\label{nevg}
\Gamma_{\rm{obs}}&=&\frac{2}{\pi}~\left<\frac{\varepsilon(t_{\rm E})}{t_{\rm E}}\right> ~\tau_{\rm{obs}}~T_{\rm{obs}},\nonumber\\
N_{\rm{event}}&=& \Gamma_{\rm{obs}} ~\left<N_{\star, \rm{MC}}\right>,
\end{eqnarray} 

\noindent where, the event rate is given in unit of $\rm{star}^{-1}\rm{season}^{-2}$, $T_{\rm{obs}}=72$ days. We estimate the average number of background source stars visible in the W149 filter towards the LMC and  SMC as $\left<N_{\star, \rm{LMC}}\right>=144616000$ and $\left<N_{\star, \rm{SMC}}\right>=42547600$ per square degree in the simulation. These values are calculated according to Equation (6) given in \citet{sajadian2019} and by averaging over the LMC and SMC area. We restrict the LMC area by $\alpha \in [80.9,~82.3]^{\circ}$ and $\delta \in [-68.2,~-66.8]^{\circ}$ and the SMC area by $\alpha \in [13.2,~14.6]^{\circ}$ and $\delta \in [-71.2,-69.8]^{\circ}$,which yields a $1.96~\rm{deg^{2}}$ area, similar to the \wfirst~observing area.

\noindent The observational optical depth, $\tau_{\rm{obs}}$ is the average optical depth across the MCs area and source distances at which their lensing signatures can be detected. For any given lens mass we estimate the average $\tau_{\rm{obs}}$, separately. %This parameter depends weakly on the lens mass. Because microlensing events caused by very low-mass microlenses are rather detectable for closer source stars. As a result, f
\begin{figure}
	\centering
	\includegraphics[width=0.49\textwidth]{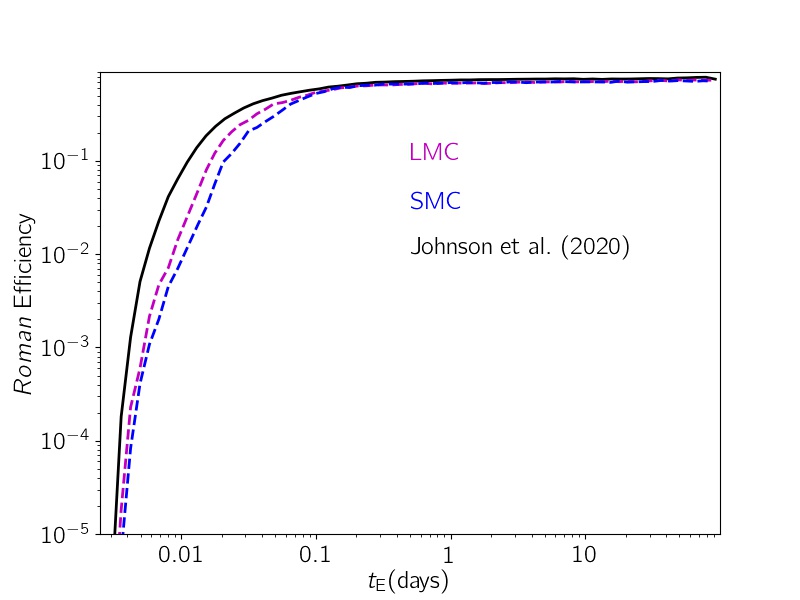}
\caption{The \wfirst~efficiencies of detecting microlensing events toward the LMC (dashed magenta line) and the SMC  (dashed blue line) versus the Einstein crossing time that were resulted from the simulations in this work. The \wfirst~efficiency for detecting microlensing events toward the Galactic bulge (represented in \citet{Penney2020}) is shown by solid black line.}\label{effiplot}
\end{figure}
\begin{figure}
\centering
\includegraphics[width=0.49\textwidth]{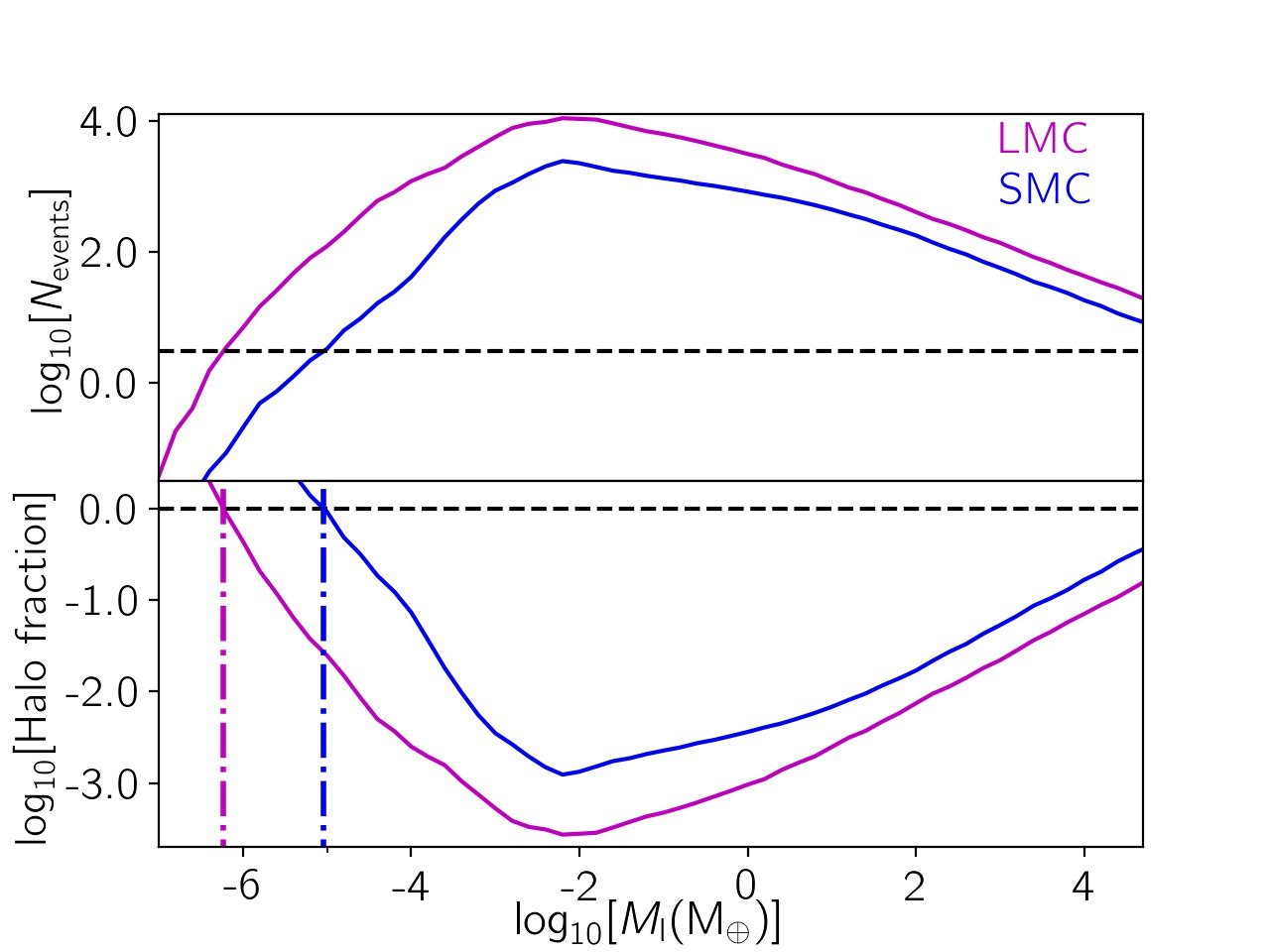}
\caption{Top panel: expected number of microlensing events that \wfirst~potentially detects by observing the LMC (magenta curve) and SMC (blue curve) during one $72$-day season per square degree. In this simulation, we choose the lens mass from a log-uniform step function and assume that the given FFP population makes up the entirety of the Galactic, LMC and SMC halos. Bottom panel: the potential $95\%$ confidence level upper limit on the exclusion of halo fractions versus the mass of FFPs. The dashed black horizontal line shows that \wfirst~potentially excludes FFPs in halos down to $5.9 \times 10^{-7} M_{\oplus}$ and $9.0 \times 10^{-6}M_{\oplus}$, if it detects no LMC and SMC events, respectively.}\label{nevent}
\end{figure}

\noindent In Equation \ref{nevg}, $\varepsilon(t_{\rm E})$ is the \wfirst~efficiency versus the Einstein crossing time which is the ratio of detectable events to total simulated events with resolvable source stars at the baseline and with the given timescale. In the simulation, we  determine these \wfirst~efficiencies for detecting the LMC and SMC microlensing events and are plotted in Figure \ref{effiplot} by dashed magenta and blue curves. In this figure, we show the \wfirst~efficiency for detecting microlensing events toward the Galactic bulge (for observations during one season) by solid black line that was offered in Figure (6) of \citet{Penney2020}.

The expected number of events, $N_{\rm{events}}$, is plotted versus the lens mass in the top panel of Figure \ref{nevent}. We assume that FFPs with the given mass (specified on the horizontal axis of the plot) form all DM halos (with $100\%$ contribution). Under this assumption, the maximum number of the MC events that \wfirst~can detect are caused by FFPs with mass $M_{\rm l} \simeq 0.006M_{\oplus}$. 

\noindent The bottom panel of Figure \ref{nevent} shows the potential $95\%$ confidence level upper limit on exclusion of halo fractions versus mass of FFPs. The black dashed horizontal line shows that \wfirst~potentially excludes FFPs (to build the entire DM halos) down to $5.9 \times 10^{-7} M_{\oplus}$ and $9.0 \times 10^{-6}M_{\oplus}$ if no LMC and SMC events are detected.  The first limit (due to the LMC observations by \wfirst) is below the current limits. For instance, the recent findings from dense-cadence observations of the Andromeda galaxy (M31) with the Subaru Hyper Suprime-Cam have probed MACHOs with  $M>10^{-11}M_{\odot} (\simeq 3 \times 10^{-6}M_{\oplus})$ to form the Galactic DM halo \citep{Niikura2019}. In addition, \citet{Griest2014} discovered no microlensing event in the first $2$-year data of the Kepler mission and excluded MACHOs with $M>6.7 \times 10^{-4}M_{\oplus}$ that make up the entirety of the Galactic DM halo.

%%%%%%%%%%%%%%%%%%%%%%%%%%%%%%%%%%%%%%%%%%%%%%%%%%%%%%%

\begin{table*}
\centering
\caption{The average values of physical and statistical parameters of microlensing events discernible by the \wfirst~telescope towards the LMC. We simulate detectable microlensing events caused by the lens objects with the discrete mass values, i.e., $M_{\rm l}=0.01,~0.1,~1,~10,~100,~1000,~10000~\rm{M}_{\oplus}$. Three rows are provided for each mass value, the first and second of which contain the average parameters of the events (I,~II). The third row contains the parameters of total events. We assume that the average number of background source stars visible in the W149 filter towards the LMC centre is $\left<N_{\star,\rm{LMC}}\right>=144616000$ per square degree, and FFPs make $5\%$ of the DM halos.}\label{table3}
\begin{tabular}{cccccccccc}\toprule[1.5pt]
&$\log_{10}[\left<M_{\rm l}\right>]$  & $\log_{10}[\left<R_{\rm E}\right>]$ & $\left<\rm{t_{\rm E}}\right>$ & $\log_{10}[\left<\rho_{\star}\right>]$ & $\log_{10}[\left<\rho_{\rm l}\right>]$ &   $\Gamma_{\rm{obs}}(10^{6})$& $\tau_{\rm{obs}}(10^{7})$ &  $N_{\rm{events}}$ &  $\epsilon[\%]$ \\
	&$(\rm{M}_{\oplus})$ & $(\rm{au})$  & $(\rm{days})$  &    &   & $(\rm{star}^{-1}~\rm{season}^{-1})$  && $(\rm{deg}^{-2} \rm{season}^{-1})$ & \\	
\toprule[1.5pt]
\multicolumn{10}{c}{$\rm{Moon-like} ~\rm{Planets}$}\\
$\rm{I}$ & $-2.0$ & $-3.33$ & $0.051$ & $-0.65$ & $-1.55$ & $10.20$ & $0.34$ & $1475.6$ & $24.3$\\
$\rm{II}$ & $-2.0$ & $-2.92$ & $0.074$ & $0.47$ & $-1.98$ & $54.03$ & $1.73$ & $7813.8$ & $2.6$\\
$\rm{Total}$ & $-2.0$ & $-3.26$ & $0.054$ & $-0.27$ & $-1.58$ & $15.19$ & $0.50$ & $2197.0$ & $12.3$\\
\hline
\multicolumn{10}{c}{$\rm{Mars-like} ~\rm{Planets} $}\\
$\rm{I}$ & $-1.0$ & $-2.87$ & $0.100$ & $-1.28$ & $-1.72$ & $9.22$ & $0.33$ & $1333.2$ & $56.1$\\
$\rm{II}$ & $-1.0$ & $-2.75$ & $0.122$ & $0.38$ & $-1.83$ & $17.52$ & $0.68$ & $2534.2$ & $39.1$\\
$\rm{Total}$ & $-1.0$ & $-2.81$ & $0.111$ & $0.07$ & $-1.77$ & $13.32$ & $0.50$ & $1926.5$ & $46.5$\\
\hline
\multicolumn{10}{c}{$\rm{Earth-like} ~\rm{Planets} $}\\
$\rm{I}$ & $0.0$ & $-2.39$ & $0.283$ & $-1.90$ & $-1.91$ & $5.76$ & $0.34$ & $833.7$ & $71.2$\\
$\rm{II}$ & $0.0$ & $-2.36$ & $0.300$ & $0.01$ & $-1.91$ & $8.67$ & $0.54$ & $1253.8$ & $67.7$\\
$\rm{Total}$ & $0.0$ & $-2.37$ & $0.292$ & $-0.26$ & $-1.91$ & $7.36$ & $0.45$ & $1064.8$ & $69.2$\\
\hline
\multicolumn{10}{c}{$\rm{Uranus-like} ~\rm{Planets} $}\\
$\rm{I}$ & $1.0$ & $-1.89$ & $0.897$ & $-2.39$ & $-1.89$ & $2.37$ & $0.33$ & $343.0$ & $78.5$\\
$\rm{II}$ & $1.0$ & $-1.87$ & $0.897$ & $-0.42$ & $-1.90$ & $3.52$ & $0.54$ & $508.4$ & $71.8$\\
$\rm{Total}$ & $1.0$ & $-1.88$ & $0.897$ & $-0.71$ & $-1.89$ & $2.99$ & $0.44$ & $432.3$ & $74.9$\\
\hline
\multicolumn{10}{c}{$\rm{Saturn-like} ~\rm{Planets} $}\\
$\rm{I}$ & $2.0$ & $-1.39$ & $3.043$ & $-2.82$ & $-1.79$ & $0.82$ & $0.33$ & $118.8$ & $79.2$\\
$\rm{II}$ & $2.0$ & $-1.37$ & $2.729$ & $-0.94$ & $-1.81$ & $1.23$ & $0.58$ & $177.9$ & $77.1$\\
$\rm{Total}$ & $2.0$ & $-1.38$ & $2.871$ & $-1.20$ & $-1.80$ & $1.07$ & $0.46$ & $154.6$ & $78.0$\\
\hline
\multicolumn{10}{c}{$\rm{Jupiter-like} ~\rm{Planets} $}\\
$\rm{I}$ & $3.0$ & $-0.90$ & $8.707$ & $-3.45$ & $-2.26$ & $0.27$ & $0.34$ & $39.0$ & $81.7$\\
$\rm{II}$ & $3.0$ & $-0.86$ & $9.044$ & $-1.46$ & $-2.28$ & $0.37$ & $0.54$ & $54.1$ & $76.2$\\
$\rm{Total}$ & $3.0$ & $-0.88$ & $8.886$ & $-1.73$ & $-2.27$ & $0.33$ & $0.44$ & $47.8$ & $78.7$\\
\hline
\multicolumn{10}{c}{$\rm{Brown} ~\rm{Dwarfs} $}\\
$\rm{I}$ & $4.0$ & $-0.40$ & $25.889$ & $-3.92$ & $-2.81$ & $0.09$ & $0.33$ & $12.7$ & $81.6$\\
$\rm{II}$ & $4.0$ & $-0.38$ & $27.288$ & $-1.90$ & $-2.82$ & $0.12$ & $0.54$ & $17.6$ & $78.6$\\
$\rm{Total}$ & $4.0$ & $-0.39$ & $26.622$ & $-2.17$ & $-2.82$ & $0.11$ & $0.44$ & $15.5$ & $80.0$\\
\hline
\end{tabular}
\end{table*}
%%%%%%%%%%%I%%%%%%%%%%%%%%%%%%%%%%%%%%%%%%%%%%%%%%%%%%%%%%%%%%%%%%%%%%%%%

We estimate the expected number of events by assuming that $5\%$ halos are formed from FFPs, which is consistent with the previous observations \citep{Tisserand2007, Niikura2019}. Hence, we repeat the simulations by considering this assumption. In Tables \ref{table3} and \ref{table4}, we summarize the details of these simulations done for some special values of the lens's mass, towards the LMC and SMC, respectively. The discrete values of the mass of microlenses (mentioned in their second columns) are $M_{\rm l}=0.01,~0.1,~1,~10,~100,~1000,~10000~M_{\oplus}$, which correspond to the Moon, Mars, Earth, Uranus, Saturn, Jupiter-like planets and brown dwarfs, respectively. However, for each given mass, three rows are provided: the first two contain average parameters of events (I,~II), respectively, and the third row contains average parameters of total events (as labeled by 'Total'). 

The \wfirst~detection efficiency, i.e., the ratio of detectable events to total simulated events with resolvable source stars at the baseline, $\epsilon[\%]$, improves from $10\%$ to $80\%$ with increasing the lens mass from $0.01M_{\oplus}$ to $10^{4}M_{\oplus}$, as given in the last column of the tables. 
%Nevertheless, the number of detectable events is not an additive function versus the lens mass. Because, by increasing the lens mass the duration of events enhances which decreases the factor $\left<\varepsilon(t_{\rm E}) /t_{\rm E}\right>$ in the event rate formula,  \citep[see, e.g., Eq. (5) of ][]{sajadian2019}. 

\noindent For the microlensing towards the LMC (Table \ref{table3}), although the events (I) have smaller finite-source effects, they are on average shorter. The first factor improves detection efficiency, whereas the second factor (short time scale) decreases it. In general, FFPs can be found in the events (I) with more efficiency than in the events (II) in the LMC observations. Towards the SMC, the self-lensing events have higher detection efficiencies. 

\noindent The efficiency for detecting Moon-like planets towards the LMC is $12\%$ which is higher than that towards the SMC. The LMC and SMC efficiencies for detecting microlensing due to Earth-mass microlenses are $69$,~$67\%$. Generally, the LMC efficiency for detecting FFPs and brown dwarfs is larger than the SMC one. Because the SMC source stars are fainter and, as a result, the photometric uncertainties for observing data points are higher.  

By making continuous observations towards the LMC, the \wfirst~telescope can detect $2197$ Moon-like,  $1065$ Earth-like, $155$ Saturn-like, and $48$ Jupiter-like planets in a $72$-day season per square degree. During one season per square degree towards the SMC, this telescope has the potential to detect $1687$ Moon-like, $654$ Earth-like, $148$ Saturn-like, and $44$ Jupiter-like planets.

\noindent It is possible to detect brown dwarfs using these observations, and this telescope will potentially confirm the $\sim16$ brown dwarfs per square degree per observing season. 

When the simulation results in the LMC and SMC directions are compared, the \wfirst~telescope can detect more microlensing events due to FFP towards the LMC than the SMC, because the number of the LMC background sources is around three times greater than the SMC one, despite the fact that the SMC optical depth is larger than the LMC one.  

\begin{table*}
\centering
\caption{Same as Table \ref{table3} but for the SMC. We assume that the average number of background source stars visible in the W149 filter toward the SMC is $\left<N_{\star,\rm{SMC}}\right>=42547600$ per square degree, and FFPs make $5\%$ of the DM halos.}\label{table4}
\begin{tabular}{cccccccccc}\toprule[1.5pt]
&$\log_{10}[\left<M_{\rm l}\right>]$  & $\log_{10}[\left<R_{\rm E}\right>]$ & $\left<\rm{t_{\rm E}}\right>$ & $\log_{10}[\left<\rho_{\star}\right>]$ & $\log_{10}[\left<\rho_{\rm l}\right>]$ &   $\Gamma_{\rm{obs}}(10^{6})$&  $\tau_{\rm{obs}}(10^{7})$& $N_{\rm{events}}$ &  $\epsilon[\%]$ \\
&$(\rm{M}_{\oplus})$ & $(\rm{au})$  & $(\rm{days})$  &    &   & $(\rm{star}^{-1}~\rm{season}^{-1})$ & & $(\rm{deg}^{-2} \rm{season}^{-1})$ &  \\	
\toprule[1.5pt]
$\rm{I}$ & $-2.0$ & $-3.07$ & $0.033$ & $-0.18$ & $-1.82$ & $11.66$ & $0.66$ & $496.2$ & $3.0$\\
$\rm{II}$ & $-2.0$ & $-2.95$ & $0.081$ & $0.62$ & $-1.97$ & $45.24$ & $1.89$ & $1924.8$ & $9.6$\\
$\rm{Total}$ & $-2.0$ & $-2.96$ & $0.074$ & $0.57$ & $-1.95$ & $39.64$ & $1.72$ & $1686.7$ & $7.3$\\
\hline
$\rm{I}$ & $-1.0$ & $-2.69$ & $0.036$ & $-1.02$ & $-1.89$ & $12.51$ & $0.63$ & $532.3$ & $22.8$\\
$\rm{II}$ & $-1.0$ & $-2.58$ & $0.140$ & $0.26$ & $-2.03$ & $27.39$ & $1.29$ & $1165.5$ & $63.9$\\
$\rm{Total}$ & $-1.0$ & $-2.59$ & $0.123$ & $0.19$ & $-2.01$ & $24.91$ & $1.18$ & $1059.8$ & $50.0$ \\
\hline
$\rm{I}$ & $0.0$ & $-2.31$ & $0.080$ & $-1.74$ & $-1.95$ & $12.87$ & $0.54$ & $547.8$ & $57.5$\\
$\rm{II}$ & $0.0$ & $-2.11$ & $0.414$ & $-0.15$ & $-2.19$ & $13.83$ & $1.22$ & $588.6$ & $70.9$\\
$\rm{Total}$ & $0.0$ & $-2.16$ & $0.314$ & $-0.30$ & $-2.10$ & $15.38$ & $1.02$ & $654.3$ & $66.5$\\
\hline
$\rm{I}$ & $1.0$ & $-1.82$ & $0.223$ & $-2.33$ & $-1.93$ & $10.66$ & $0.64$ & $453.6$ & $70.9$\\
$\rm{II}$ & $1.0$ & $-1.59$ & $1.384$ & $-0.65$ & $-2.21$ & $5.43$ & $1.31$ & $231.0$ & $78.0$\\
$\rm{Total}$ & $1.0$ & $-1.65$ & $1.010$ & $-0.82$ & $-2.10$ & $8.96$ & $1.10$ & $381.2$ & $75.6$\\
\hline
$\rm{I}$ & $2.0$ & $-1.33$ & $0.688$ & $-2.78$ & $-1.78$ & $4.57$ & $0.62$ & $194.3$ & $76.6$\\
$\rm{II}$ & $2.0$ & $-1.09$ & $4.194$ & $-1.22$ & $-2.13$ & $1.55$ & $1.21$ & $66.1$ & $79.1$\\
$\rm{Total}$ & $2.0$ & $-1.17$ & $2.923$ & $-1.41$ & $-1.97$ & $3.48$ & $0.99$ & $148.1$ & $78.2$\\
\hline
$\rm{I}$ & $3.0$ & $-0.82$ & $2.528$ & $-3.30$ & $-2.30$ & $1.36$ & $0.61$ & $58.0$ & $77.6$\\
$\rm{II}$ & $3.0$ & $-0.60$ & $13.438$ & $-1.68$ & $-2.60$ & $0.51$ & $1.20$ & $21.6$ & $78.8$\\
$\rm{Total}$ & $3.0$ & $-0.66$ & $9.828$ & $-1.85$ & $-2.48$ & $1.03$ & $1.00$ & $43.8$ & $78.4$\\
\hline
$\rm{I}$ & $4.0$ & $-0.34$ & $6.534$ & $-3.86$ & $-2.81$ & $0.49$ & $0.61$ & $20.8$ & $78.4$\\
$\rm{II}$ & $4.0$ & $-0.09$ & $43.698$ & $-2.21$ & $-3.14$ & $0.18$ & $1.29$ & $7.8$ & $77.6$\\
$\rm{Total}$ & $4.0$ & $-0.17$ & $30.194$ & $-2.40$ & $-2.99$ & $0.40$ & $1.04$ & $16.9$ & $77.9$\\
\hline
\end{tabular}
\end{table*}

%%%%%%%%%%%I%%%%%%%%%%%%%%%%%%%%%%%%%%%%%%%%%%%%%%%%%%%%%%%%%%%%%%%%%%%%%%
\section{Conclusions}\label{five}
Here, we studied the properties and statistics of detectable short-duration microlensing events due to FFPs through microlensing observations of the Magellanic Clouds.

The relative lens-source velocities in microlensing events towards the LMC are, on average, $\sim40-50 \rm{km s^{-1}}$ lower than those towards the Galactic bulge, which are $\sim 120-150\rm{km s^{-1}}$. Furthermore, the lens distance $D_{\rm l}$ to the source distance $D_{\rm s}$ ratio is either $x_{\rm{rel}} \lesssim 0.15$ (when lenses are inside the Galactic halo, which has small projected source radii) or $x_{\rm{rel}} \gtrsim 0.85$ (when lenses are inside the LMC which result large projected source radii). 

\noindent Accordingly, an FFP in the Galactic halo creates longer microlensing events with smaller projected source radii toward the LMC in comparison to similar events towards the Galactic bulge. A similar FFP makes shorter microlensing events toward the SMC because their relative lens-source velocities are larger, $\sim 140-150 \rm{km s^{-1}}$. Self-lensing events towards the MCs suffer from severe finite-source effects, although they are even longer.  

\noindent By simulating microlensing events in the direction of MCs from FFPs, we estimated the efficiencies to reach SNRs higher than $50$ for these events. These efficiencies are $\sim 0.5\%$ (for events with lenses inside the Galactic halo) and $\sim 0.2\%$ (for self-lensing events) via ground-based optical surveys (in the $VIK$ bands). Through space-based near-infrared surveys, the efficiencies are higher and $\sim 4\%$ (with lenses inside in the Galactic halo) and $\sim 3\%$ (self-lensing events). Hence, the highest efficiency for capturing microlensing signatures due to FFPs in the Galactic halo is achievable by observing the LMC through a space-based near-infrared survey. The \wfirst~mission is a good candidate for such observations. 

Considering the dense-cadence microlensing survey by \wfirst~telescope, we have done more realistic simulations of short-duration microlensing events due to FFPs. We assumed that \wfirst~would observe each Magellanic Cloud during one observing season with the $15.16$min cadence and estimated the number of detectable FFPs. The \wfirst~efficiencies (versus mass) for detecting FFPs increases from $\sim 10\%$ to $\sim80\%$ by increasing their masses from $0.01M_{\oplus}$ to $10^{4}M_{\oplus}$ (in events with resolvable source stars at the baseline). 

By assuming a log-uniform step function for FFPs mass and considering $100\%$ as their contribution in DM halos, we estimated the expected number of events as plotted in Figure \ref{nevent}, and consequently, $95\%$ confidence level upper limit on the exclusion of FFPs dark matter (versus mass) which potentially will be achievable in the \wfirst~observations (if it detects no events). We concluded that \wfirst~potentially extends the mass range of detectable FFPs in halos down to $5.9 \times 10^{-7}M_{\oplus}$ (and $9.0 \times 10^{-6}M_{\oplus}$) with continuous observations over one $72$-day season towards the LMC (and SMC).  This limit is lower than the recent limit on MACHOs fraction in the Galactic DM halo by \citet{Niikura2019}. 

By assuming a log-uniform step function for FFPs mass and considering $5\%$ as their contribution in DM halos, we concluded that \wfirst~potentially discovers $2200$-$1700$, $1060$-$650$ and $155$-$148$ short-duration microlensing events due to the Moon, Earth, and Saturn-mass FFPs towards the LMC,~SMC per observing season per square degree, respectively. It is possible to detect brown dwarfs using these observations and this telescope will discover $\sim 16$ brown dwarfs per observing season per square degree. The maximum numbers of FFPs that \wfirst~can detect are due to those with a mass of $0.01M_{\oplus}$. 

\section*{Acknowledgements}
I thank K.~Dobashi and Y.~C.~Joshi  for kindly providing the LMC and SMC extinction maps. I acknowledge M.~Penny and S.~Johnson for offering the efficiency plot for detecting short-duration microlensing events by \wfirst. I especially thank the referee for the helpful comments and good suggestions. The author thanks the Department of Physics, Chungbuk National University, and especially C.~Han for hospitality. 

\section*{DATA AVAILABILITY}
The data underlying this article will be shared on reasonable request to the corresponding author.
\bibliographystyle{mnras}
\bibliography{paper_ref}
\appendix
\section{Mass densities in the Magellanic Clouds}\label{append1}
As explained in section \ref{three}, for simulating MCs structures, we need two coordinate systems. The first one is the observer coordinate system, $(x,~y,~z)$. Its $z-$axis is towards the observer, $x-$axis is anti-parallel to the right ascension axis and $y-$axis is parallel with the declination axis. The second one, $(x',~y',~z')$, is the coordinate system of the Magellanic Clouds. Its $x'$ and $y'-$ axes are parallel with the semi-major and semi-minor axes of the Magellanic Clouds' disc, respectively. The centres of both coordinate systems are on the Magellanic Clouds centre. 

\noindent The position of any given point in the first coordinate system is converted to the second one as following \citep[see, e.g., Appendix of ][]{Weinberg2001}:
\begin{eqnarray}\label{convert}
x' &=& x \sin \theta -  y  \cos \theta, \nonumber\\
y' &=&x \cos \theta  \cos i + y \sin \theta \cos i -z\sin i,\nonumber\\
z' &=&x \cos \theta \sin i+ y \sin \theta \sin i + z\cos i,
\end{eqnarray}
We note that $\theta$ and $i$ are the projection angles, whose values are given in subsections \ref{lmc} and \ref{smc}. The LMC and SMC mass densities in their coordinate systems, i.e., $(x',~y',~z')$, which are used in the work are in the following.  

\textbf{The LMC mass densities}:  The LMC structure has been well studied in many references \citep[see, e.g., ][]{VanderMarel2001a, VanderMarel2001b, VanDerMarel2002}. The mass density of the LMC disc, which is used in this paper, in the second coordinate system is given by \citep{Kim2000}:
\begin{eqnarray}
\rho_{d, \rm{L}}= \frac{M_{\rm{d}, \rm{L}} }{4 \pi z_{\rm d} R_{\rm d}^{2}} \exp(-\frac{R'}{R_{\rm d}}) \exp(- \frac{\abs{z'}}{z_{\rm d}} ), 
\end{eqnarray}

\noindent where $M_{\rm{d}, \rm{L}}=2.6 \times 10^{9}~\rm{M}_{\odot}$ is the mass of the LMC disc, $z_{\rm d}=0.3$ kpc and $R_{\rm d}=1.8$ kpc.  We note that $R'=\sqrt{x'^{2} + y'^{2}}$. 

\noindent The mass density profile of the LMC bar can be given by \citep{Kim2000}: 

\begin{eqnarray}
\rho_{b, \rm{L}}= \frac{M_{\rm b}}{(2 \pi)^{3/2} z_{\rm b} y_{\rm b} x_{\rm b}}\exp(\frac{-1}{2} [(\frac{x'}{x_{\rm b}})^{2}  + (\frac{y'}{y_{\rm b}})^{2}  + (\frac{z'}{z_{\rm b}})^{2} ]),
\end{eqnarray}

\noindent where $z_{\rm b}= y_{\rm b}=0.44$ kpc and $x_{\rm b}=1.2$ kpc. The LMC bar's mass is $M_{\rm b}\simeq 0.15 M_{\rm d}$.

\noindent For the halo of the LMC and SMC, we consider a mass density with spherical symmetry as the following \citep{Kim2000}:  
\begin{eqnarray}\label{halos}
\rho_{h}= \frac{ \rho_{0,\rm h} }{1+  r^{2}/R^{2}_{\rm h} },
\end{eqnarray}

\noindent where  $r^{2}=x^{2} +  y^{2} +z^{2}$.  For the LMC, $R_{\rm h}=2~$kpc and $\rho_{0, \rm h}=1.76 \times 10^{7}~M_{\odot} \rm{kpc}^{-3}$. We assume that the radius of the LMC halo is $15~$kpc. For the SMC, the parameters are $R_{\rm h}=1.5~$kpc, $\rho_{0, \rm h}= 9.3 \times 10^{6}~M_{\odot}\rm{kpc}^{-3}$, which is corresponding to the total dynamic mass around $1.7 \times 10^{9}M_{\odot}$.  We consider a cut radius for the SMC halo at $8.5~$kpc. %In the simulation, we assume around $20\%$ of the halo can be in the form of MACHOs (in this paper, FFPs) and will act as microlenses. 

\textbf{The SMC mass densities}:  The small Magellanic Cloud has two structures, a disc and a surrounding halo (its mass density is given in the previous paragraph).  For its disc, we consider two components of young and old stellar populations (YS,~OS, respectively).  Their mass densities are given in the following \citep{Calchi2013}: 

\begin{eqnarray}
\rho_{\rm {d, YS}}= \rho_{1} \exp(-\frac{1}{2} [(\frac{x'}{X_{\rm d}})^{2}  + (\frac{y'}{Y_{\rm d}})^{2}  + (\frac{z'}{Z_{\rm d} })^{2}]),
\end{eqnarray}

\noindent where, $\rho_{1}= 6\times 10^{6}M_{\odot}\rm{kpc}^{-3}$, $X_{\rm d},~Y_{\rm d},~Z_{\rm d}$ are $0.8,~3.5,~1.3$ kpc, respectively. The mass density of the old stellar population is given by:  

\begin{eqnarray}
\rho_{\rm d, \rm{OS}}= \rho_{2} \exp(-\sqrt{ (\frac{x'}{X_{\rm d}})^{2} +(\frac{y'}{Y_{\rm d}})^{2}})\exp (-\frac{1}{2} \frac{z'^{2}}{Z_{\rm d}}) 
\end{eqnarray}
\noindent where, $\rho_{2}= 2.7\times10^{7}M_{\odot}\rm{kpc}^{-3}$ and $X_{\rm d},~Y_{\rm d},~Z_{\rm d}$ are $0.8,~1.2,~2.1$ kpc, respectively.
\end{document}